\documentclass[useAMS,usenatbib]{mn2e}

\usepackage[a4paper,total={6.90in, 9.6in},top=85pt,left=42pt,asymmetric=true]{geometry} 

\usepackage[T1]{fontenc}
\usepackage{aecompl}

\usepackage{mathptmx} 
\usepackage{threeparttable,threeparttablex,multirow,booktabs,longtable} 
\usepackage{afterpage} 
\usepackage{moresize} 
\usepackage{graphicx} 
\usepackage{amsmath,amssymb} 
\usepackage{color}
\usepackage[ampersand]{easylist}
\usepackage[large,bf,singlelinecheck=false]{caption} 
\usepackage{float} 

\title[Inclinations and Titius-Bode Predictions of Kepler Systems]{Using the Inclinations of Kepler Systems to Prioritize New Titius-Bode-Based 
Exoplanet Predictions}

\author[T. Bovaird, C. H. Lineweaver and S. K. Jacobsen]{T. Bovaird$^{1,2}$\thanks{E-mail: timothy.bovaird@anu.edu.au}, C. H. Lineweaver$^{1,2,3}$ and S. K. Jacobsen$^{4}$ \\ 
$^{1}$Research School of Astronomy and Astrophysics, Australian National University, Canberra, ACT 2611, Australia \\ 
$^{2}$Planetary Science Institute, Australian National University \\ 
$^{3}$Research School of Earth Sciences,Australian National University\\
$^{4}$Niels Bohr Institute, University of Copenhagen, Copenhagen, Denmark}

\begin{document} 
\pagerange{\pageref{firstpage}--\pageref{lastpage}} \pubyear{2015}

\maketitle 
\label{firstpage}

\begin{abstract} 
{We analyze a sample of multiple-exoplanet systems which contain at least 3 transiting planets detected by the \emph{Kepler} mission (`Kepler multiples'). 
We use a \emph{generalized} Titius-Bode relation to predict the periods of 228 additional planets in 151 of these Kepler multiples. 
These Titius-Bode-based predictions suggest that there are, on average, 
$2 \pm 1$
planets in the habitable zone of each star.
We estimate the inclination of the invariable plane for each system and
prioritize our planet predictions by their geometric probability to transit. 
We highlight a short list of 77 predicted planets in 40 systems with a high geometric probability to transit, resulting in 
an expected detection rate of $\sim 15$ per cent, $\sim3$ times higher than the detection rate of our previous Titius-Bode-based predictions.
 } 
\end{abstract}

\section{INTRODUCTION} 
\label{sec:intro}

The Titius-Bode (TB) relation's successful prediction of the period of Uranus was the main motivation that led to the search for another planet between Mars and Jupiter, e.g. \cite{Jaki1972}.   
This search led to the discovery of the asteroid Ceres and the rest of the asteroid belt. 
The TB relation may also provide useful hints about the periods of as-yet-undetected planets around other stars.
In \cite{Bovaird2013} (hereafter, BL13) we used a \emph{generalized} TB relation
to analyze 68 multi-planet systems with four or more detected exoplanets. 
We predicted the existence of 141 new exoplanets in these 68 systems. 
\cite{Huang2014} (hereafter, HB14) performed an extensive search in the Kepler data for 97 of our predicted planets in 56 systems. 
This resulted in the
confirmation of 5 of our predictions. 
(Fig. \ref{fig:confirmed_predictions} and Table \ref{tab:haungdetectiontable}).

In this paper we perform an improved TB analysis on a larger sample of Kepler multiple-planet systems\footnote{
Accessed November 4th, 2014: http://exoplanetarchive.ipac.caltech.edu/cgi-bin/TblView/nph-tblView?app=ExoTbls\&config=cumulative}
to make new exoplanet orbital period predictions.
We use the expected coplanarity of multiple-planet systems to estimate the most likely inclination of the invariable plane of each system.
We then prioritize our original and new TB-based predictions according to their geometric probability of transiting.
Comparison of our original predictions with the HB14 confirmations shows that restricting our predictions to those with a high geometric probability to transit 
should increase the detection rate by a factor of $\sim 3$ (Fig. \ref{fig:Ptrans_histo}). 

As in BL13, our sample includes all Kepler multi-planet systems with four or more exoplanets, 
but to these we add three-planet systems if the orbital periods of the system's planets adhere better to the TB relation than the Solar System (Eq. 4 of BL13). 
Using these criteria we add 77 three-planet systems to the 74 systems with four or more planets. 
We have excluded 3 systems:  KOI-284, KOI-2248 and KOI-3444 
because of concerns about false positives due to adjacent-planet period ratios close to 1 and close binary hosts \citep{Lissauer2011a,Fabrycky2014,Lillo2014}. 
We have also excluded the three-planet system KOI-593, since the period of KOI-593.03 was recently revised, excluding the system from our three-planet sample. 
Thus, we analyze 151 Kepler multiples, with each system containing 3, 4, 5 or 6 planets.
%
%
\begin{figure*} 
	\includegraphics[width=1.0\textwidth]{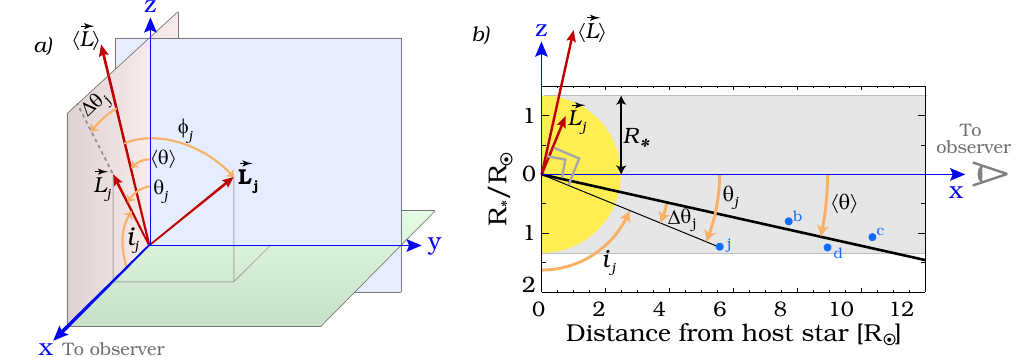} 
\caption{\emph{Panel a):} Our coordinate system for transiting exoplanets.  The $x$-axis points towards the observer. 
$\vec{\bf{L}_j}$ is the 3-D angular momentum of the $j$th planet, and is perpendicular to the orbital plane of the $j$th planet.
$\vec{\langle L\rangle}$ is the sum of the angular momenta of all detected planets (Eq. \ref{eq:L}) and is perpendicular to the invariable plane of the system.
We have chosen the coordinate system without loss of generality such that $\vec{\langle L\rangle}$ has no component in the y direction.
 $\phi_{j}$ is the angle between $\vec{\bf{L}_j}$ and $\vec{\langle L\rangle}$.
 Let $\vec{L_j}$ be the projection of $\vec{\bf{L}_{j}}$ onto the $x-z$ plane.
 $\Delta \theta_j$ is the angle between $\vec{\langle L\rangle}$ and $\vec{L_j}$.
 $\langle \theta \rangle$ is the angle between the $z$ axis and $\vec{\langle L\rangle}$.
 $i_j$ is the inclination of the planet (Eq. \ref{eq:i}).
 $\theta_j = 90 - i_j$ and is the angle between the $z$ axis and $\vec{L_j}$ such that 
 $\theta_j = \langle \theta \rangle +  \Delta \theta_j $ (Eq. \ref{eq:deltatheta}).
 \emph{Panel b)} shows the $x-z$ plane of Panel a) with the $y$ axis pointing into the paper. The observer is to the right.
The grey shaded region represents the `transit region' where the centre of a planet will transit its host star as seen by the 
observer (impact parameter $b\leq1$, see Eq. \ref{eq:b}). 
Four planets, b, c, d and $j$, are represented by blue dots. The intersection of the orbital plane of the $j$th planet with the $x-z$ plane is shown (thin black line). 
All angles shown in Panel a) (with the exception of $\phi_j$) are also shown in Panel b.
The thick line is the intersection of the invariable plane of the system with the $x-z$ plane.
Because we are only dealing with systems with multiple transiting planets, all these angles are typically less than a few degrees but are exaggerated here for clarity. } 
	\label{fig:plspac_inc_single}
\end{figure*}

\subsection{Coplanarity of exoplanet systems} 
\label{sec:exoplanet_coplanarity}

Planets in the Solar System and in exoplanetary systems are believed to form from protoplanetary disks (e.g. \cite{Winn2014}). 
The inclinations of the 8 planets of our Solar System to the invariable plane are (in order from Mercury to Neptune) 
$6.3^\circ{}$, $2.2^\circ{}$, $1.6^\circ{}$, $1.7^\circ{}$, $0.3^\circ{}$, $0.9^\circ{}$, $1.0^\circ{}$, $0.7^\circ{}$ \citep{Souami2012}. 
Jupiter and Saturn contribute $\sim 86$ per cent of the total planetary angular momentum and thus the angles between their orbital planes and the invariable plane are small:  $0.3^\circ{}$ and $0.9^\circ{}$ respectively.

\begin{figure} 
	\includegraphics[width=0.47\textwidth]{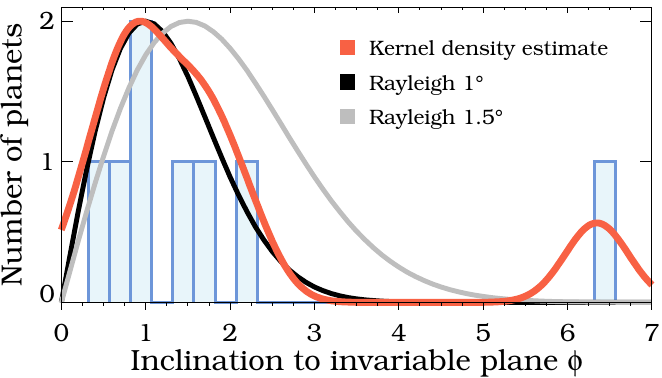} 
\caption{The coplanarity of planets in the Solar System relative to the invariable plane. With the exception of Mercury, the angles between the orbital planes of the planets and the invariable plane are well represented by a Rayleigh distribution with a mode of $\sim1^\circ$.} 
\label{fig:solsys_rayleigh} 
\end{figure}

In a given multiple-planet system, the distribution of mutual inclinations between the orbital planes of planets is well described by a Rayleigh distribution  \citep{Lissauer2011a,Fang2012a,Figueira2012,Fabrycky2014,Ballard2014}. 
For the ensemble of Kepler multi-planet systems, 
the mode of the Rayleigh distribution of \emph{mutual inclinations} ($\phi_j-\phi_i$) is typically $\sim1^{\circ} - 3^{\circ}$
(Appendix \ref{app:coordinates} \& Table \ref{tab:prev_studies}).
Thus, Kepler multiple-planet systems are highly coplanar. 
The Solar System is similarly coplanar.  
For example, the mode of the best fit Rayleigh distribution of the planet inclinations $(\phi_j)$ 
\emph{relative to the invariable plane} in the Solar System is $\sim 1^{\circ}$ (see Fig. \ref{fig:plspac_inc_single} and \ref{fig:solsys_rayleigh})\footnote{See Appendix \ref{app:invplane_calculation} 
for an explanation of why the distribution of mutual inclinations is on average a factor of $\sqrt{2}$ wider than the distribution of the angles
$\phi_j$ in Fig. \ref{fig:plspac_inc_single},
between the invariable plane of the system and the orbital planes of the planets.}.  

The angle $\Delta \theta_{j}$ is a Gaussian distributed variable with a mean of $0$ (centered around $\langle \theta \rangle$) 
and standard deviation $\sigma_{\Delta \theta}$.
Based on previous analyses (Table \ref{tab:prev_studies}), 
we assume the typical value $\sigma_{\Delta \theta} =1.5^{\circ}$.
We use this angle to determine the probability of detecting additional transiting planets in each system. 

Estimates of the inclination of a transiting planet come from the impact parameter $b$ which is the projected distance between
the center of the planet at mid-transit and the center of the star, in units of the star's radius. 
\begin{equation} 
	\label{eq:b} 
	b= \frac{a}{R_*} \cos{i} 
\end{equation}
where $R_*$ is the radius of the star and $a$ is the semi-major axis of the planet.
For edge-on systems, typically $ 85^{\circ} < i < 95^{\circ}$.
However, since we are unable to determine whether $b$ is in the positive $z$ direction or the negative $z$ direction (Fig.\ref{fig:plspac_inc_single}b and Fig. \ref{fig:var_illustration}),
we are unable to determine whether $i$ is greater than or less than $90^{\circ}$. 
By convention, for transiting planets the sign of $b$ is taken as positive and thus the corresponding $i$ values from Eq. \ref{eq:b} are taken as $ i \le 90^{\circ}$.

The impact parameter is also a function of four transit light curve observables \citep{Seager2003}; the period $P$, the transit depth $\Delta F$, 
the total transit duration $t_T$, and the total transit duration minus the ingress and egress times $t_F$ (the duration where the light curve is flat for a source uniform across its disk).
Thus the impact parameter can be written,
\begin{equation} 
	\label{eq:b2} 
	b= f(P, \Delta F, t_T, t_F).
\end{equation}
Eliminating $b$ from Eqs. (\ref{eq:b}) and (\ref{eq:b2}) yields the inclination $i$ as a function of observables,
\begin{equation} 
	\label{eq:i} 
	i=\cos^{-1}\left[\frac{R_*}{a} f(P, \Delta F, t_T, t_F) \right] 
\end{equation}
From Eq. ~\ref{eq:b} we can see that for an impact parameter $b=0$ (a transit through the center of the star), 
we obtain $i=90^\circ$; an `edge-on' transit.

The convention $i \le  90^{\circ}$ is unproblematic when only a single planet is found to transit a star but raises an issue when multiple exoplanets transit the same star, 
since the degree of coplanarity depends on whether the actual values of $i_j$ ($j=1,2,..N$ where $N$ is the number of planets in the system) are greater than or less than $90^{\circ}$.
For example, the actual values of $i_j$ in a given system could be all $> 90^{\circ}$, all $ <90^{\circ}$ or some in-between combination. 
Although we do not know the signs of $\theta_j = 90^{\circ} - i_j$ for individual planets, we can estimate the inclination
of the invariable plane for each system, by calculating all possible permutations of the $\theta_j$ values for each system (see Appendix~\ref{app:permutations}).
In this estimation, we use the plausible assumption that the coplanarity of a system should not depend on the inclination 
of the invariable plane relative to the observer.

\subsection{The probability of additional transiting exoplanets} 
\label{sec:acrit} 
We wish to develop a measure of the likelihood of additional transiting planets in our sample of Kepler multi-planet
systems. The more edge-on a planetary system is to an observer on Earth, the greater the probability of a planet transiting at larger periods.
Similarly, a larger stellar radius leads to a higher probability of additional transiting planets (although with a reduced detection efficiency). 
We quantify these tendencies under the assumption that Kepler multiples have a Gaussian opening angle $\sigma_{\Delta\theta}=1.5^\circ$ 
around the invariable plane,
and we introduce the variable, $a_\text{crit}$.
Planets with a semi-major axis greater than $a_\text{crit}$ have less than a $50 \%$ geometric probability of transiting.
More specifically, $a_\text{crit}$ is defined as the semi-major axis where $P_{\text{trans}}(a_\text{crit}) = 0.5$ (Eq. \ref{eq:ptrans}) for a given system.

\begin{figure} 
	\includegraphics[width=0.47\textwidth]{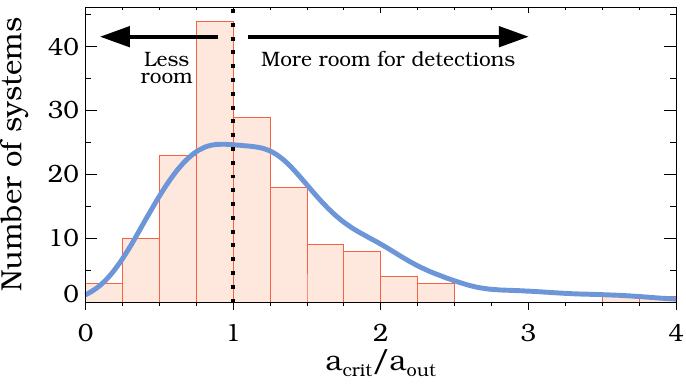} 
\caption{Histogram of the $a_\text{crit}/a_\text{out}$ values for our sample of Kepler multiples. The distribution peaks
for a ratio just below 1. Approximately half of the systems lie to the right of 1. For these systems, if there are planets with semi-major axes $a$  such that $ a_\text{out} < a < a_\text{crit}$
then the geometric probability of them transiting is greater than $50\%$.
Note that this does not account for the detectability of these planets (e.g. they could be too small to detect). 
The majority of the
predicted planets that are insertions ($a < a_\text{out}$) 
have geometric transit probabilities greater than 50 per cent, when $a_\text{crit}/a_\text{out}<1$.
Systems on the right have more room for detections, and in general, predicted planets in these systems have higher values of $P_{\text{trans}}$.
The blue curve is the expected distribution of our sample of Kepler multiples if they all have planets at TB-predicted semi-major axes
extrapolated out to $ \sim 4 \times a_\text{crit}$. 
The blue curve is consistent with the observed distribution, indicating that our $a_\text{crit}/a_\text{out}$ distribution is consistent with the system in our sample containing more planets than have been detected.
} 
\label{fig:acrit_aout_histo} 
\end{figure}

In a given system, a useful ratio for estimating the amount of semi-major axis space where additional transiting planets are more likely, 
is $a_\text{crit}/a_\text{out}$,
where $a_\text{out}$ is the semi-major axis of the detected planet in the system which is the furthest from the host star.
The larger $a_\text{crit}/a_\text{out}$, the larger the semi-major axis range for additional transiting planets beyond the outermost detected planet. 
Values for this ratio less than 1 mean that the outermost detected planet is beyond the calculated $a_\text{crit}$ value, and imply that additional transiting planets 
beyond the outermost detected planet are less likely. 
Figure~\ref{fig:acrit_aout_histo} shows the $a_\text{crit}/a_\text{out}$ distribution for all systems in our sample.
The fact that this distribution is roughly symmetric around  $a_\text{crit}/a_\text{out} \sim 1$ strongly suggests that the 
outermost {\it transiting} planets in Kepler systems are due to the inclination of the system to the observer, and
are not really the outermost planets. 


In Section~\ref{sec:followup} we discuss the follow-up that has been done on our BL13 planet detections. 
In Section~\ref{sec:transitprob} 
we show that the $\sim 5\%$ follow-up detection rate of HB14 is consistent with selection effects and the existence of the 
predicted planets.
In Section~\ref{sec:predictions} we extend and upgrade the TB relation developed in BL13 and predict the periods of 
undetected planets in our updated sample. We then prioritize these predictions based on their geometric probability to transit and 
emphasize for further follow-up a subset of predictions with high transit probabilities. 
We also use TB predictions to estimate the average number of planets in the circumstellar habitable zone.
In Section~\ref{sec:periodratios} we discuss how our predicted planet insertions affect the period ratios of adjacent planets
and explore how period ratios are tightly dispersed around the mean period ratio within each system.
In Section~\ref{sec:conclusion} we summarize our results.

\begin{figure*} 
\includegraphics[width=1.0\textwidth]{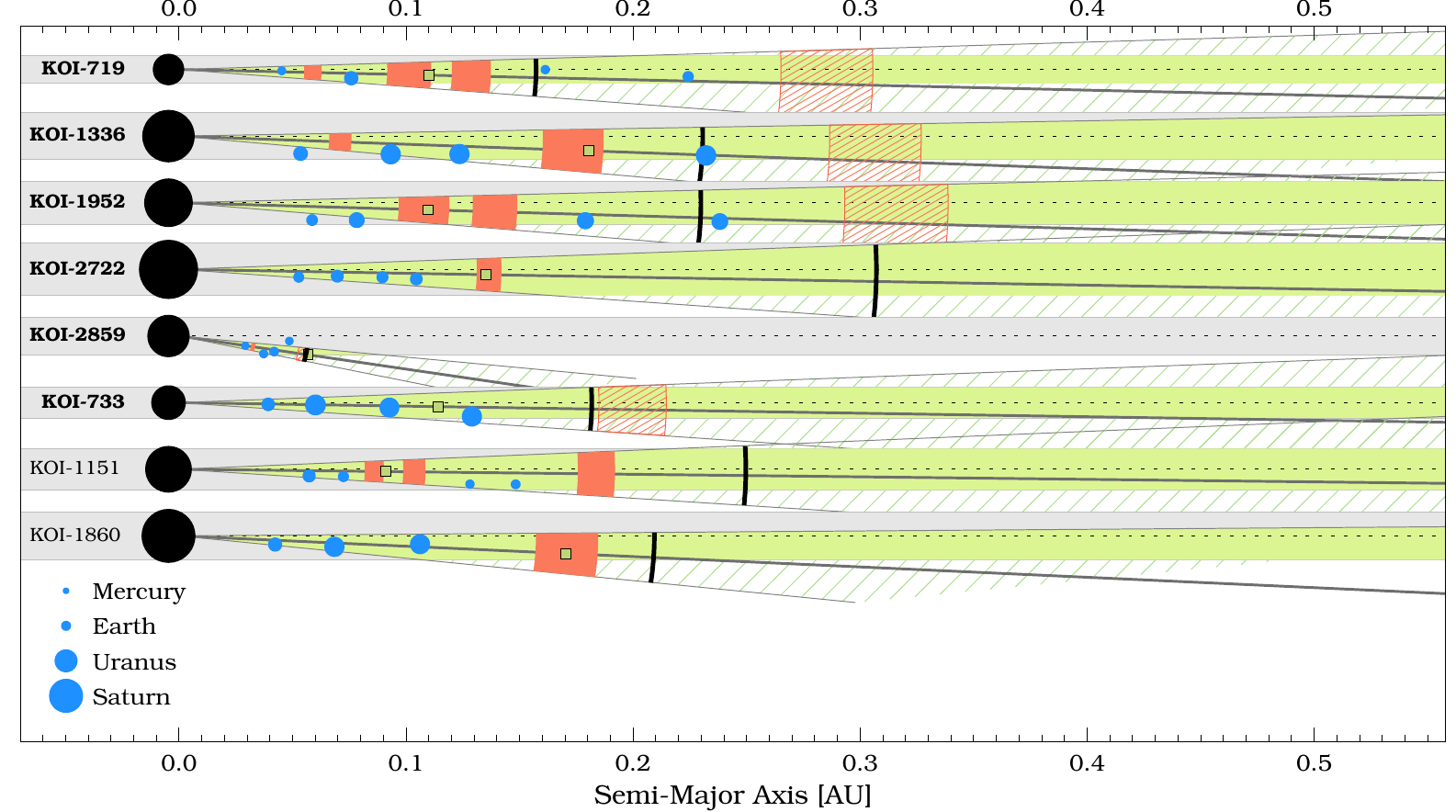}
\caption{Exoplanet systems where an additional candidate was detected after a TB relation prediction was made (see Table~\ref{tab:haungdetectiontable}). 
The systems are shown in descending order of $a_\text{crit}/a_\text{out}$.
Previously known planets are shown as blue circles. 
The predictions of BL13 and their uncertainties are shown by the red filled rectangles if the $P_\text{trans}$ value of 
the predicted planet is $\ge 0.55$
(Eq.~\ref{eq:ptrans} and Fig.~\ref{fig:Ptrans_histo}), or by red hatched rectangles otherwise. 
The new candidate planets are shown as green squares. 
The critical semi-major axis $a_\text{crit}$ (Section~\ref{sec:acrit}), beyond which $P_{\text{trans}}(a_\text{crit}) < 0.5$, 
 is shown by a solid black arc. 
The uncertainties (width of red rectangles) in this figure and Table~\ref{tab:haungdetectiontable}, are slightly wider than Figure~\ref{fig:confirmed_predictions_numins2}
 due to excessive rounding of predicted period uncertainties for some systems in BL13.} 
\label{fig:confirmed_predictions} 
\end{figure*}
\begin{figure*}
	\includegraphics[width=1.0\textwidth]{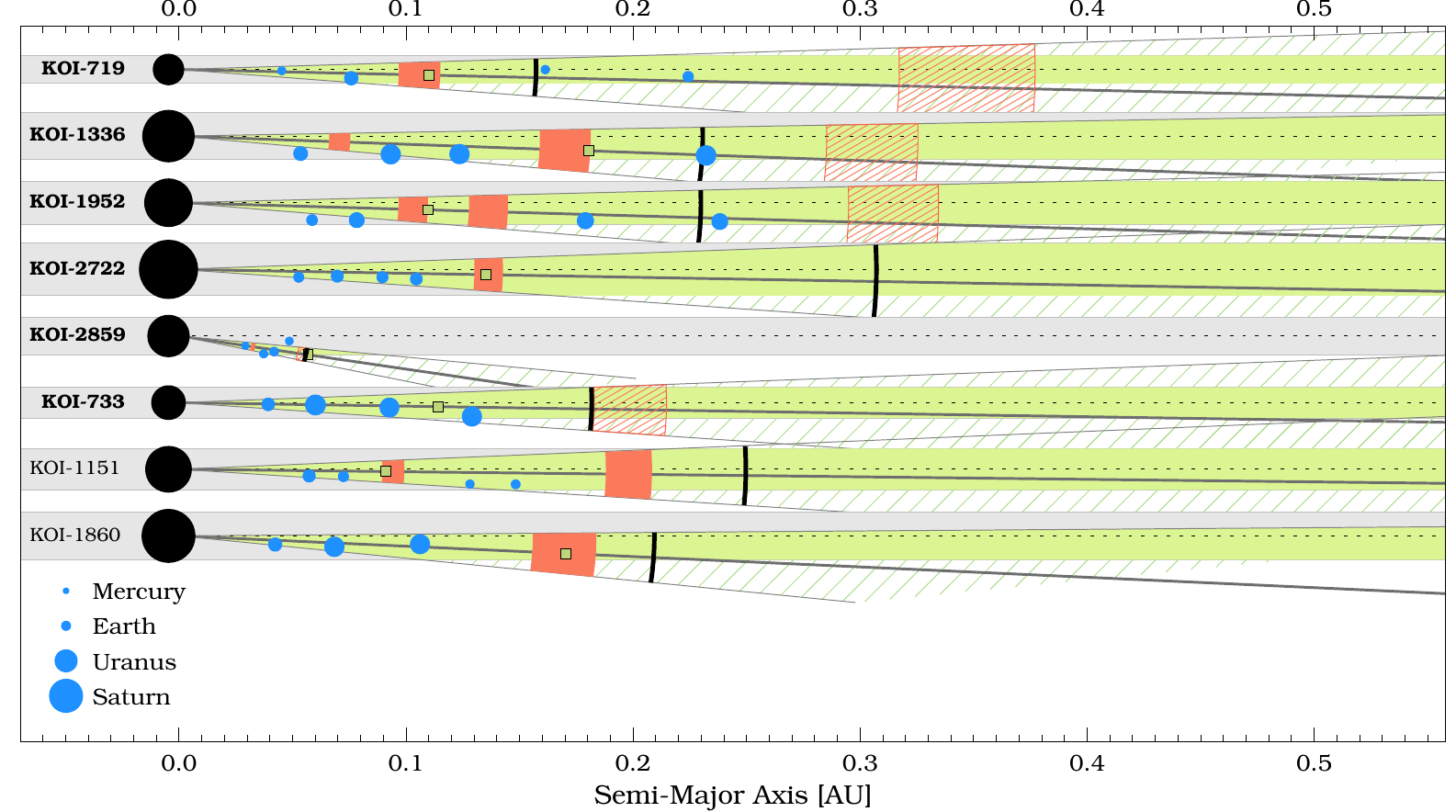}
\caption{Same as Fig.~\ref{fig:confirmed_predictions} except here our TB predictions are based on $\gamma$ with $n_{ins}^{2}$ (Eq.~\ref{eq:gamma}) 
rather than on the $\gamma$ with $n_{ins}$ of Eq. 5 of BL13.
Comparing Fig.~\ref{fig:confirmed_predictions} with this figure, in the  KOI-719 system the number of predicted planets goes from 4 to 2, 
while in KOI-1151 the number of predicted planets goes from 3 to 2. In both cases the detected planet is more centrally located in the predicted region.
}
	\label{fig:confirmed_predictions_numins2}
\end{figure*}

\begin{table} 
\begin{ThreePartTable}
\caption{\normalsize Systems with candidate detections by HB14 (in bold)
plus KOI-1151 \protect\citep{Petigura2013b} and KOI-1860 $^b$, after planet predictions were made by BL13.
}
\setlength{\tabcolsep}{3.5pt} 
\begin{tabular}{lcccc} \hline 
System & Predicted & Detected & Predicted & Detected \\ 
& Period (days) & Period (days) & Radius ($R_\oplus$) & Radius ($R_\oplus$)\\ 
\hline
{\bf KOI-719} & $14\pm2$ & 15.77 & $\le 0.7$ & 0.42 \\ 
{\bf KOI-1336} & $26\pm3$ & 27.51 & $\le 2.4$ & 1.04 \\ 
{\bf KOI-1952} & $13\pm2$ & 13.27 & $\le 1.5$ & 0.85 \\ 
{\bf KOI-2722$^a$} & $16.8\pm1.0$ & 16.53 & $\le 1.6$ & 1.16 \\ 
{\bf KOI-2859} & $5.2\pm0.3$ & 5.43 & $\le 0.8$ & 0.76\\ 
{\bf KOI-733} & N/A & 15.11 & N/A & 3.0\\ 
KOI-1151$^a$ & $9.6\pm0.7$ & 10.43 & $\le 0.8$ & 0.7\\
KOI-1860$^b$ & $25\pm3$ & 24.84 & $\le 2.7$ & 1.46\\

\hline 
\end{tabular}
\begin{tablenotes}
\item[a] Predicted by preprint of BL13 (draft uploaded 11 Apr 2013: $\text{http://arxiv.org/pdf/1304.3341v1.pdf}$), 
detected planet reported by Kepler archive and included in analysis of BL13.
\item[b] October 2014 Kepler Archive update, during the drafting of this paper.
\end{tablenotes}
\label{tab:haungdetectiontable} 
\end{ThreePartTable}
\end{table}

\section{FOLLOW-UP OF BL13 PREDICTIONS} 
\label{sec:followup} 
BL13 used the approximately even logarithmic spacing of exoplanet systems to make predictions for the periods of 
$117$ additional candidate planets in $60$ Kepler detected systems, 
and $24$ additional predictions in $8$ systems detected via radial velocity(7) and direct imaging(1), which we do not consider here. 
NASA Exoplanet Archive data updates, confirmed our prediction 
of KOI-2722.05 (Table \ref{tab:haungdetectiontable}).

HB14 used the planet predictions made in BL13 to search for 97 planets in the light curves of 56 Kepler systems. 
Within these 56 systems, BL13 predicted the period and maximum radius:  the largest
radius which would have evaded detection, based on the lowest signal-to-noise of the detected planets in the same system. 
Predicted planets were searched for using the Kepler Quarter 1 to Quarter 15 long cadence
light curves, giving a baseline exceeding 1000 days. Once the transits of the already known planets were detected and removed, 
transit signals were visually inspected around the predicted periods.

\begin{figure}
\includegraphics[width=0.47\textwidth]{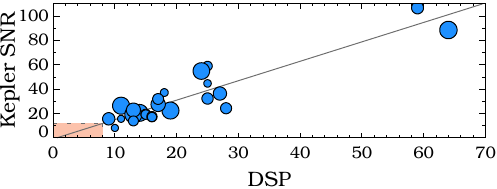}
\caption{Reported dip significance parameter (DSP) from Table 1 of \protect\cite{Huang2014} for previously known exoplanets in the 
five systems with a new detection. A linear trend can be seen for the DSP and the signal-to-noise ratio as reported by the 
Kepler team \protect\citep{Christiansen2012}. 
HB14 required planet candidates to have a DSP $>8$ to survive their vetting process.
The sizes of the blue dots correspond to the same planetary radii representation used to make Figs. 
\ref {fig:planetspacing_inclination1}, \ref{fig:planetspacing_inclination2},
\ref{fig:planetspacing_inclination3} \& \ref{fig:allsystems_hz1}.}
\label{fig:dspvssnr}
\end{figure}

Of the 97 predicted planets searched for by HB14, 5 candidates were detected within $\sim$ one-sigma of the predicted periods
(5 planets of the 6 planets in bold in Table~\ref{tab:haungdetectiontable}, see also Fig. \ref{fig:confirmed_predictions}). 
Notably all new planet candidates have Earth-like or lower planetary radii.
One additional candidate was detected in KOI-733 which is incompatible with the predictions of BL13. 
This candidate is unique in that it should 
have been detected previously, based on the signal-to-noise of the other detected planets in KOI-733. 
In Table \ref{tab:haungdetectiontable}, the detected radii are less than the maximum predicted radii in each case. 
The new candidate in KOI-733 has a period of 15.11 days and a radius of 3 $R_\oplus$. 
At this period, the maximum radius to evade detection should have been 2.2 $R_\oplus$.
With the possible exception of KOI-1336 where a dip significance parameter ($DSP$, \cite{Kovacs2005}) was not reported, 
all detected candidates have a $DSP$  of $\ge 8$, 
which roughly corresponds to a Kepler SNR \citep{Christiansen2012} of $\gtrsim 12$ (see Figure~\ref{fig:dspvssnr}).
 HB14 required $DSP >8$ for candidate transit signals to survive their vetting process.

\begin{figure}
\includegraphics[width=0.47\textwidth]{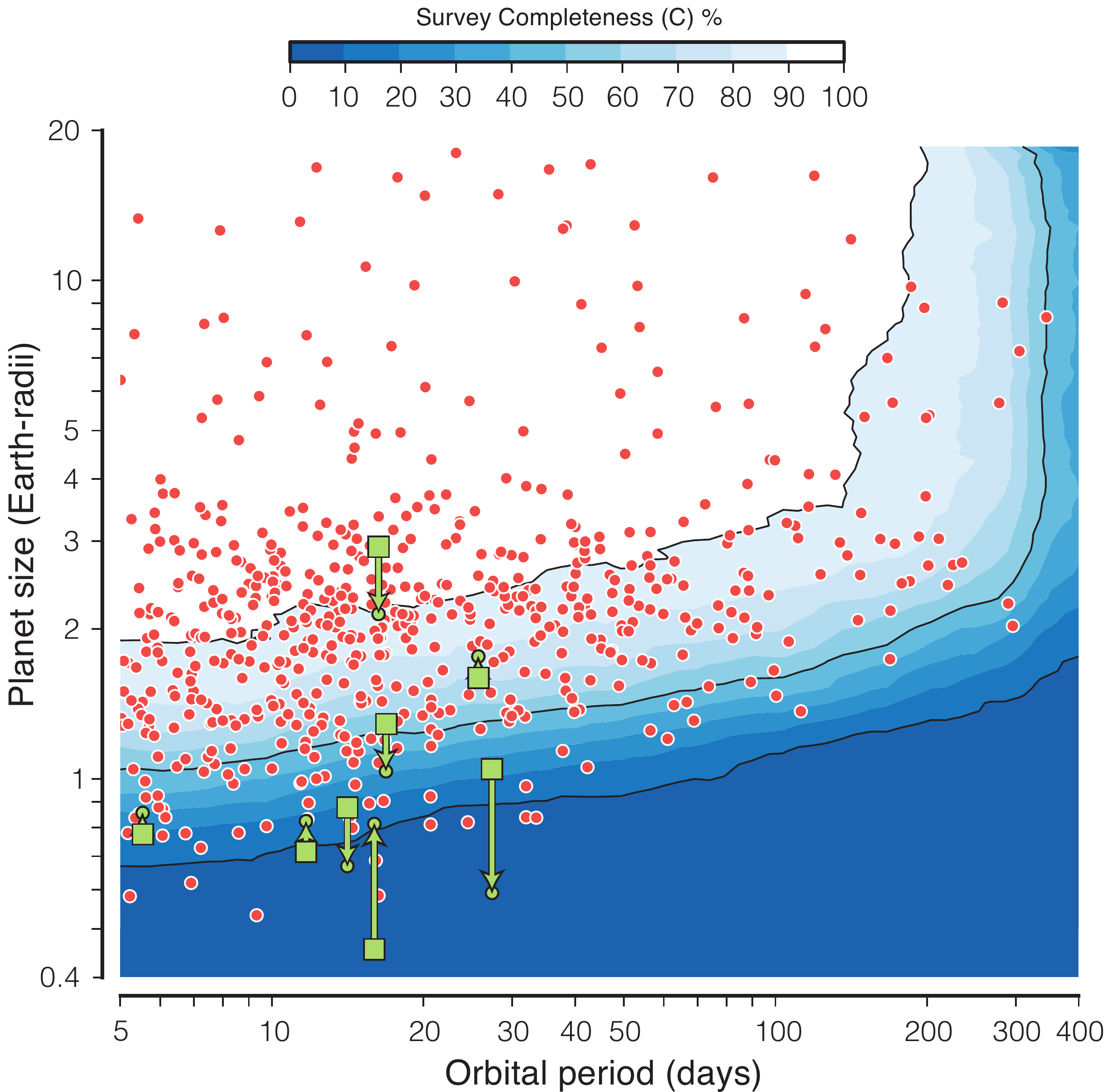}
\caption{The simulated detection completeness of the new candidate planets in the TERRA pipeline 
(modified from Figure 1 of \protect\cite{Petigura2013b}). Here we overplot as green squares, the 8 planets listed in 
Table~\ref{tab:haungdetectiontable}. The completeness curves are averaged over all stellar noise and stellar radii in 
the \protect\cite{Petigura2013b} sample (the 42,000 least noisy Kepler stars). Green circles indicate the `effective radius' of the 
new candidates, based on the noise and radius of their host star in comparison to the median of the quietest 42,000 sample. 
From the signal-to-noise, the effective radius can be calculated by 
$R_{p\text{,eff}}/R_p = (R_*/R_{*\text{,median}}) \times(CDPP/CDPP_\text{median})^{1/2}$, where $CDPP$ is 
the combined differential photometric precision defined in \protect\cite{Christiansen2012}. Taking a subset of 42,000 stars from the Kepler input catalog with the 
lowest 3-hour CDPP (approximately representative of the sample in the figure), we obtain
$CDPP_\text{median}\approx60\text{ ppm}$ and $R_{*\text{,median}}\approx1.15 \text{ }R_\odot$. 
Using the effective radius and excluding the outlier KOI-733, the mean detection completeness for the 7 candidate planets 
in Table \ref{tab:haungdetectiontable} is $\sim24\%$. 
}
\label{fig:petigura_completeness}
\end{figure}

\section{Is a $5\%$ detection rate consistent with selection effects?} 
\label{sec:transitprob}
From a sample of 97 BL13 predictions, HB14 confirmed 5.
However, based on this $\sim 5 \%$  detection rate, HB14 concluded that the predictive power of the TB relation 
used in BL13 was questionable.
Given the selection effects, how high a detection rate should one expect?
We do not expect all planet predictions to be detected. The predicted planets may have too large an inclination to transit relative to the observer.
Additionally, there is a completeness factor due to the intrinsic noise of the stars, the size of the planets, and the 
techniques for detection.
This completeness for Kepler data has been estimated for the automated lightcurve analysis pipeline TERRA \citep{Petigura2013b}. 
Fig.~\ref{fig:petigura_completeness} displays the TERRA pipeline injection/recovery completeness. 
After correcting for the radius and noise of each star, relative to the TERRA sample in Fig.~\ref{fig:petigura_completeness}, 
the planet detections in Table \ref{tab:haungdetectiontable} have an 
average detection completeness in the TERRA pipeline of $\sim24 \%$. 
That is, if all of our predictions were correct 
and if all the planets were in approximately the same region of period and radius space as the green squares in Fig. \ref{fig:petigura_completeness},
and if all of the planets transited, we would expect a detection rate of $\sim 24\%$ using the TERRA pipeline. 
It is unclear how this translates into a detection rate for a manual investigation of the lightcurves motivated by TB predictions.

We wish to determine, from coplanarity and detectability arguments, how many of our BL13 predictions we would have expected to be detected. 
An absolute number of expected detections is most limited by the poorly known
planetary radius distribution below 1 Earth radius 
\citep{Howard2012,Dressing2013,Dong2013,Petigura2013,Fressin2013,Silburt2014,Morton2014,Foreman-Mackey2014}.
Large uncertainties about the shape and amplitude of the planetary radius distribution of rocky 
planets with radii less than 1 Earth radius 
make the evaluation of TB-based exoplanet predictions difficult.
Since the TB relation predicted the asteroid belt ($M_{asteroid} < 10^{-3} M_{Earth}$) 
there seems to be no lower mass limit to the objects that the TB relation can predict.
This makes estimation of the detection efficiencies strongly dependent on assumptions about the frequency of planets at small radii.

Let the probability of detecting a planet, $P_\text{detect}$, be the product of the geometric probability to transit $P_\text{trans}$ as seen by the 
observer (Appendix \ref{app:Ptrans})
and the probability $P_\text{SNR}$ that the planetary radius is large enough to produce a signal-to-noise ratio above the detection threshold,
\begin{equation} P_\text{detect}=P_\text{trans}  \: \:P_\text{SNR}. 
\label{eq:pdetect} 
\end{equation} 

\begin{figure} 
\includegraphics[width=0.47\textwidth]{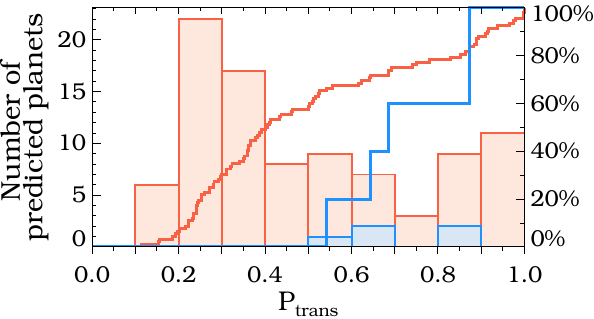} 
\caption{Histogram of $P_\text{trans}$ (Eq.~\ref{eq:ptrans}), the geometric probability of transit, for the 97 
 predicted planets from  BL13, that were followed up by HB14.
The blue histogram represents the 5 new planets 
detected by HB14 (Table~\ref{tab:haungdetectiontable}). 
As expected, the detected planets have high $P_\text{trans}$ values compared to the entire sample. 
The red and blue solid lines represent the empirical cumulative distribution function for the two distributions.
A K-S test of the two distributions yields a p-value of $1.8\times10^{-2}$.
Thus, $P_\text{trans}$ can be used to prioritize our predictions and increase their probability of detection.} 
\label{fig:Ptrans_histo} 
\end{figure}


\begin{figure} 
\includegraphics[width=0.47\textwidth]{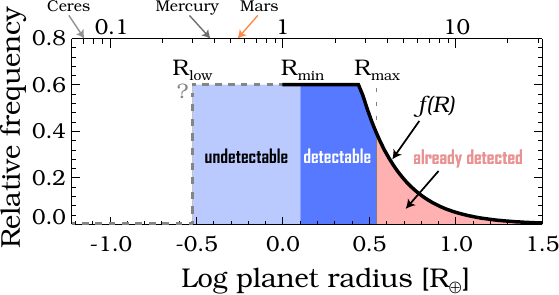} 
\caption{The assumed distribution of planetary radii described in Section \ref{sec:transitprob}. 
The distribution is poorly constrained below 1 Earth radii, indicated by the gray dashed line.
For low mass stars, the planetary radius distribution may decline below 0.7 $R_\oplus$ \citep{Dressing2013, Morton2014}. 
Alternative estimations show the planetary radius distribution continuing to increase with smaller radii 
(continuing the flat logarithmic distribution), down to 0.5 $R_\oplus$ \citep{Foreman-Mackey2014}.
For our analysis we have extrapolated the flat distribution (in log R) down to $R_\text{low}$. 
We indicate three regions for a hypothetical system at a specific predicted period. 
The `already detected' region refers to the range of planetary radii which should already have been detected, based on the lowest signal-to-noise 
ratio of the detected planets in that system. 
$R_\text{min}$ is the smallest radius which could produce a transit signal that exceeds the detection threshold, and is the boundary between the 
undetectable and detectable regions.} 
\label{fig:planetary_radius_distribution_illustration} 
\end{figure}

The geometric probability to transit, $P_\text{trans}$, is defined in Eq. \ref{eq:ptrans} and illustrated in Fig.~\ref{fig:plspac_inc_single} and 
Fig.~\ref{fig:var_illustration}.
The 5 confirmations from our previous TB predictions are found in systems with a much higher than random probability of transit (Fig. \ref{fig:Ptrans_histo}).
This is expected if our estimates of the invariable plane are reasonable.

To estimate $P_\text{SNR}$ we first estimate the probability that the radius of the planet will be large enough to detect. 
In BL13 we estimated the maximum planetary radius, $R_\text{max}$, for a hypothetical undetected planet at a given period, based on
the lowest signal-to-noise of the detected planets in the same system. We now wish to estimate a minimum radius that would be detectable, given the individual noise of each star.
We refer to this parameter as $R_\text{min}$, which is the minimum planetary radius that Kepler could detect around a given star (using a specific SNR threshold). 
For each star we used the mean $CDPP$ (combined differential photometric precision) noise from Q1-Q16. When the number of transits is not reported, we use 
the approximation $N_\text{trans}\approx T_\text{obs} f_0/P$, where $T_\text{obs}$ is the total observing time and $f_0$ is the fractional observing uptime, 
estimated at $\sim$0.92 for the Kepler mission \citep{Christiansen2012}.

\begin{figure} 
\includegraphics[width=0.47\textwidth]{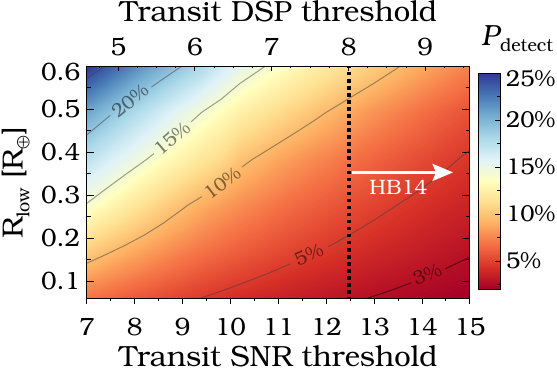}
\caption{The mean detection rate $P_\text{detect}$ (Eq. \ref{eq:pdetect}) of the BL13 predictions is dependent on the transit
signal-to-noise threshold ($SNR_{th}$ in Eq. \ref{eq:Rmin})
used in lightcurve vetting (x-axis), and on the probability of low-radii planets, 
i.e. on how far the flat logarithmic planetary radius distribution in Fig. \ref{fig:planetary_radius_distribution_illustration} 
should be extrapolated ($R_\text{low}$ on the y-axis). 
For example, in the denominator of Eq. \ref{eq:pSNR}, integrating down to a radius $R_\text{low}$ of 0.6 $R_\oplus$ and setting a 
SNR threshold of 7 (which sets $R_\text{min}$ in the numerator)  
gives an expected detection fraction of $\sim 25\%$ (blue values in the upper left  of the plot).
Integrating down to a radius of 0.2 $R_\oplus$ and having a $DSP$ threshold of 8 (converted from $SNR$ according to Fig.~\ref{fig:dspvssnr})
gives an  expected detection fraction of $\sim 5\%$ (red values in the lower right). 
} 
\label{fig:SNRRpcontour} 
\end{figure}

The probability $P_\text{SNR}$ depends on the underlying planetary radius probability density function.
We assume a density function of the form: 
\begin{equation} 
f(R)=\frac{df}{d{\log R}}= \begin{cases} k(\log R)^\alpha, &  R \ge 2.8\: R_\oplus \\ k(\log 2.8)^\alpha, & R <  2.8\: R_\oplus \end{cases} 
\end{equation} 
where $k=2.9$ and $\alpha=-1.92$ \citep{Howard2012}. 
The discontinuous distribution accounts for the approximately flat number of planets per star in logarithmic planetary radius bins for
$R \lesssim 2.8 \;R_\oplus$
\citep{Dong2013,Fressin2013,Petigura2013,Silburt2014}. 
For $R \lesssim 1.0\;R_\oplus$ the distribution is poorly constrained. For this paper, we extend the flat distribution in 
log $R$ down to a minimum radius $R_\text{low} = 0.3\; R_\oplus $. 
It is important to note that for the Solar System, the poorly constrained part of the planetary radius distribution contains 50 per cent of the planet population.
For reference the radius of Ceres, a ``planet''  predicted by the TB relation applied to our Solar System 
has a radius $R_{Ceres} = 476 \:\text{km} = 0.07 R_\oplus$.

The probability that the hypothetical planet has a radius that exceeds the SNR detection threshold is then given by 
\begin{equation} 
P_\text{SNR}=\frac{\int_{R_\text{min}}^{R_\text{max}}f(R) dR}{\int_{R_\text{low}}^{R_\text{max}}f(R) dR} 
\label{eq:pSNR} 
\end{equation} 
We do not integrate beyond $R_\text{max}$ since we expect a planet with a radius greater than $R_\text{max}$ would have already been detected. 
We define $R_\text{max}$ by,
\begin{equation}
R_\text{max} = R_\text{min SNR} \left(\frac{P_\text{predict}}{P_\text{min SNR}}\right)^{1/4},
\label{eq:Rmax}
\end{equation}
where $R_\text{min SNR}$ and $P_\text{min SNR}$ are the radius and period respectively of the detected planet with the lowest signal-to-noise in the system.
$P_\text{predict}$ is the period of the predicted planet.
$R_\text{min}$ depends on the $SNR$ in the following way: 
\begin{equation}
R_\text{min} = R_* \sqrt{SNR_\text{th}\;CDPP} \left(\frac{3 \,hrs}{n_\text{tr} \: t_\text{T}}\right)^{1/4} \: ,
\label{eq:Rmin}
\end{equation}
where $SNR_\text{th}$ is the $SNR$ threshold for a planet detection, 
$n_\text{tr}$ is the number of expected transits at the given period and $t_\text{T}$ is the transit duration in hours.
See Figure \ref{fig:planetary_radius_distribution_illustration} for an illustration of how the integrals in 
$P_\text{SNR}$ (Eq. \ref{eq:pSNR})
depend on the planet radii limits, $R_\text{min}$ and $R_\text{low}$.

While $P_\text{trans}$ is well defined, $P_\text{SNR}$ is dependent on the $SNR$ threshold chosen ($SNR_\text{th}$), the choice of $R_\text{low}$ and 
the poorly constrained shape of the planetary radius distribution below 1 Earth radius. 
This is demonstrated in Figure~\ref{fig:SNRRpcontour}, where the mean $P_\text{detect}$ from the predictions of BL13 
(for $DSP > 8$)
can vary from 
$\sim 2$ per cent to $\sim 11$ per cent.
Performing a K-S test on $P_\text{SNR}$ values (analogous to that in Figure~\ref{fig:Ptrans_histo}) indicates that the $P_\text{SNR}$ values for the subset of our BL13 
predictions that were detected, are drawn from the same $P_\text{SNR}$ distribution as all of the predicted planets. 
For this reason we use only $P_\text{trans}$, the geometric probability to transit, to prioritize our new TB relation predictions. 
We emphasize a subset of our predictions which have a $P_\text{trans}$ value $\ge$ 0.55, since all of the confirmed predictions of BL13 had a $P_\text{trans}$ value above
this threshold. 
Only $\sim 1/3$ of the entire sample have $P_\text{trans}$ values this high.
Thus, the $\sim 5$ per cent detection rate should increase by a factor of $\sim 3$ to $\sim 15$ per cent for
our new
high-$P_\text{trans}$ subset
of planet period predictions.

\section{UPDATED PLANET PREDICTIONS} \label{sec:predictions}

\subsection{Method and Inclination Prioritization}

\begin{figure*} 
\includegraphics[width=1.00\textwidth]{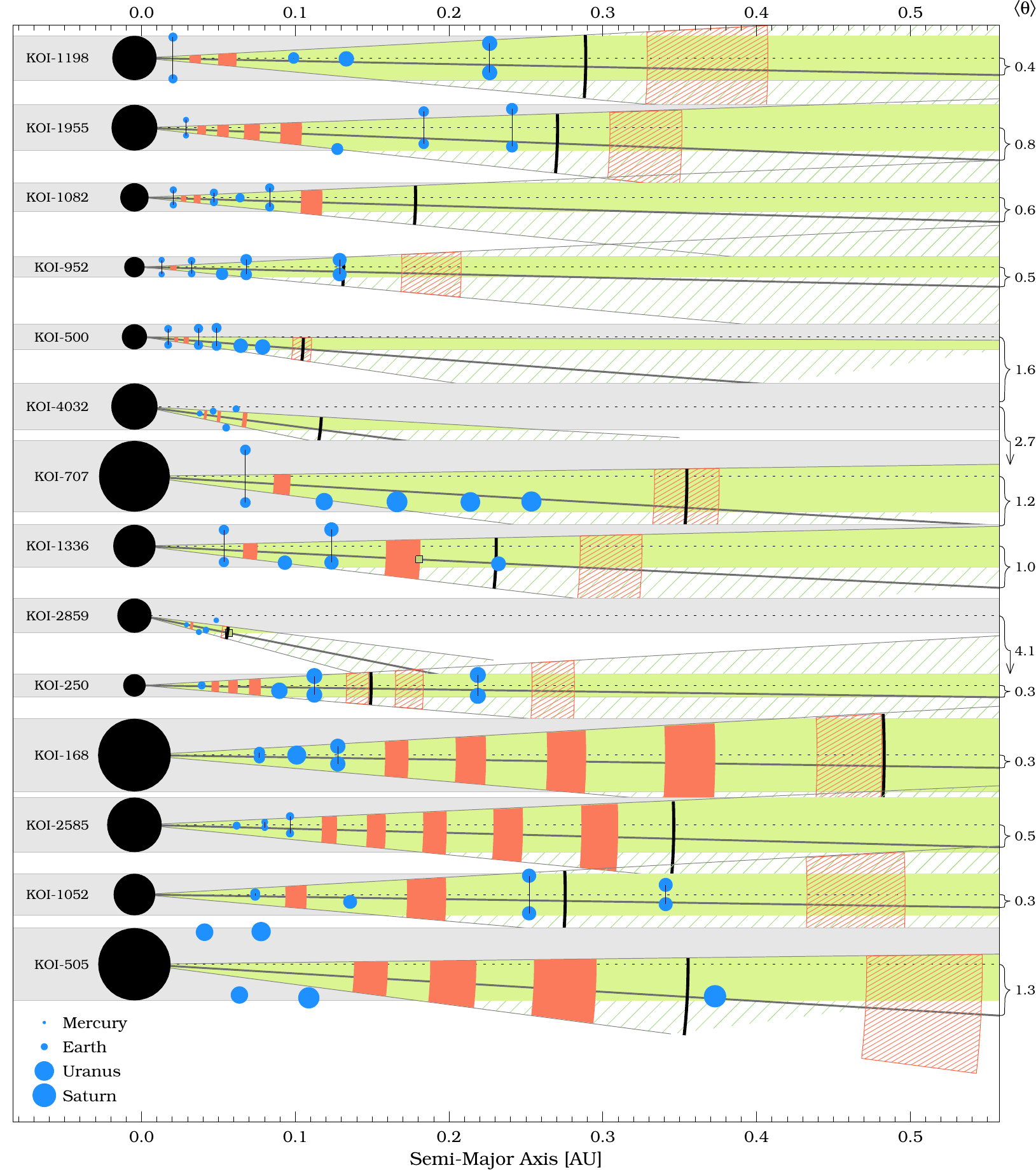} 
\caption{The architectures and invariable plane inclinations for Kepler systems in our sample which contain at least one planet 
with a geometric probability to transit $P_\text{trans}\ge0.55$. There are 40 such systems out of the 151 in our sample. 
The 14 with the highest $P_\text{trans}$ values are plotted here.
The remaining 26 are plotted in the next two figures. 
The order of the systems (from top to bottom) is determined by the highest $P_\text{trans}$ value in each system.
The thin horizontal dotted line represents the line-of-sight to Earth, i.e. where the $i$ value of a planet would be $90^\circ{}$. 
The thick grey line in each system is our estimate of the invariant plane angle, $\langle\theta\rangle$ (Appendix \ref{app:permutations}).
The value of $\langle\theta\rangle$ is given in degrees to the right of each panel (see also Fig. \ref{fig:iinvplanepdf_dispersion}).
The green wedge has an opening angle $\sigma_{\Delta\theta} = 1.5^{\circ}$ and is symmetric around the invariable plane,
but is also
limited to the grey region where a planet can be seen to transit from Earth ($b\le1$, Eq. \ref{eq:b}).
The thick black arc indicates the $a_\text{crit}$ value beyond which less than 50 per cent of planets will transit (Eq.~\ref{eq:ptrans}).
Predicted planets and their uncertainties are shown by solid red rectangles if the $P_\text{trans}$ value of the predicted planet is $\ge 0.55$, or 
by red hatched rectangles otherwise.
Thus, the 77 solid red rectangles in the 40 systems shown in 
Figs. \ref{fig:planetspacing_inclination1},\ref{fig:planetspacing_inclination2} \& \ref{fig:planetspacing_inclination3}
make up our short list of highest priority predictions (Table~\ref{tab:predictions}1).
Our estimate of the most probable inclination ambiguities in a system are represented by vertically separated pairs of blue dots, connected by a thin
black line (see Appendix \ref{app:permutations}).} 
\label{fig:planetspacing_inclination1}
\end{figure*}

\begin{figure*} 
\includegraphics[width=1.00\textwidth]{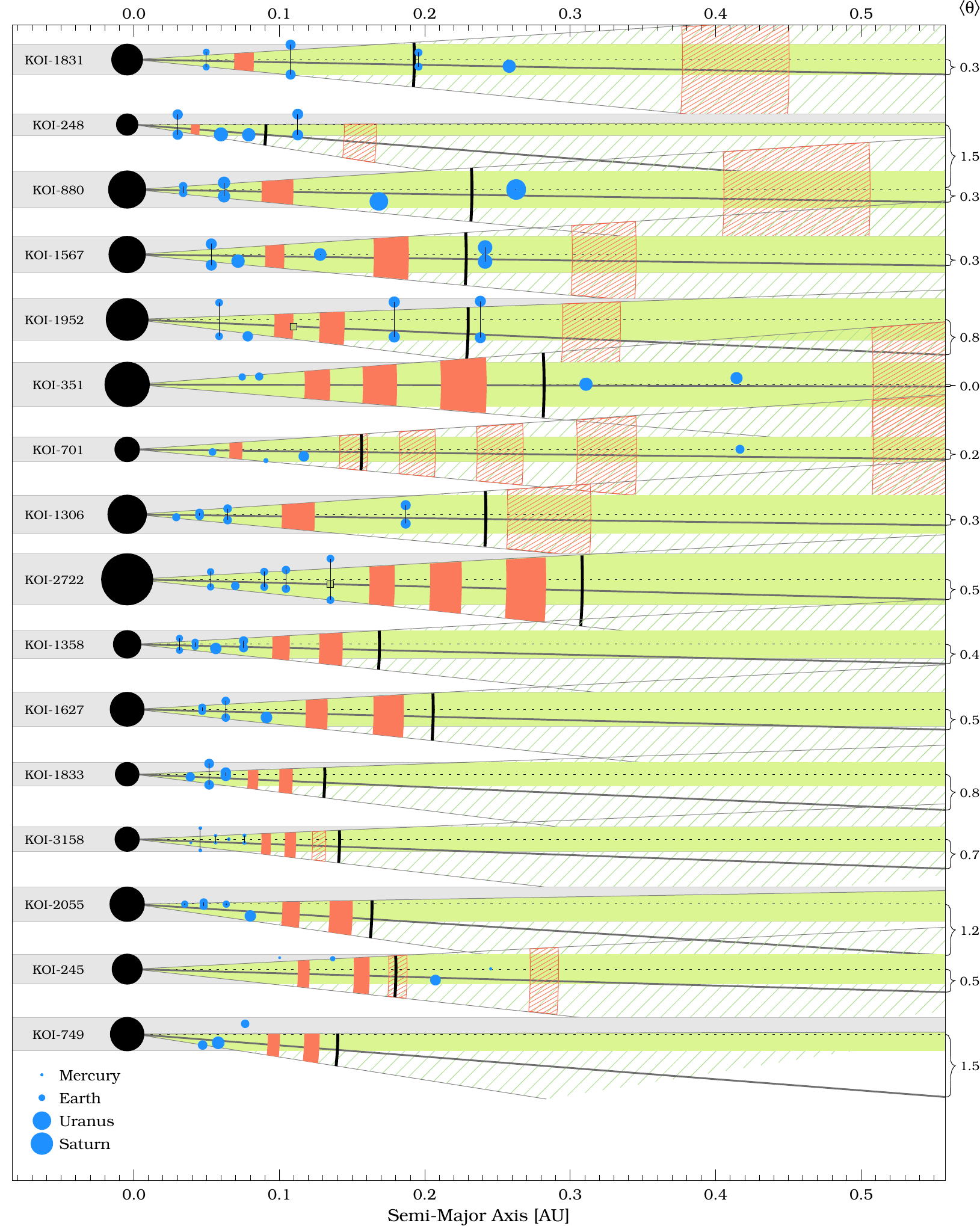}
\caption{The same as Figure~\ref{fig:planetspacing_inclination1} but for the next 16 systems in our sample which have at least 
one planet with $P_\text{trans}\ge0.55$. 
Note that some of the new detected planets from the predictions of BL13 have been included in the Kepler data archive (see Table \ref{tab:haungdetectiontable}), 
and that these planets are included in our analysis.} 
\label{fig:planetspacing_inclination2} 
\end{figure*}

\begin{figure*} 
\includegraphics[width=1.00\textwidth]{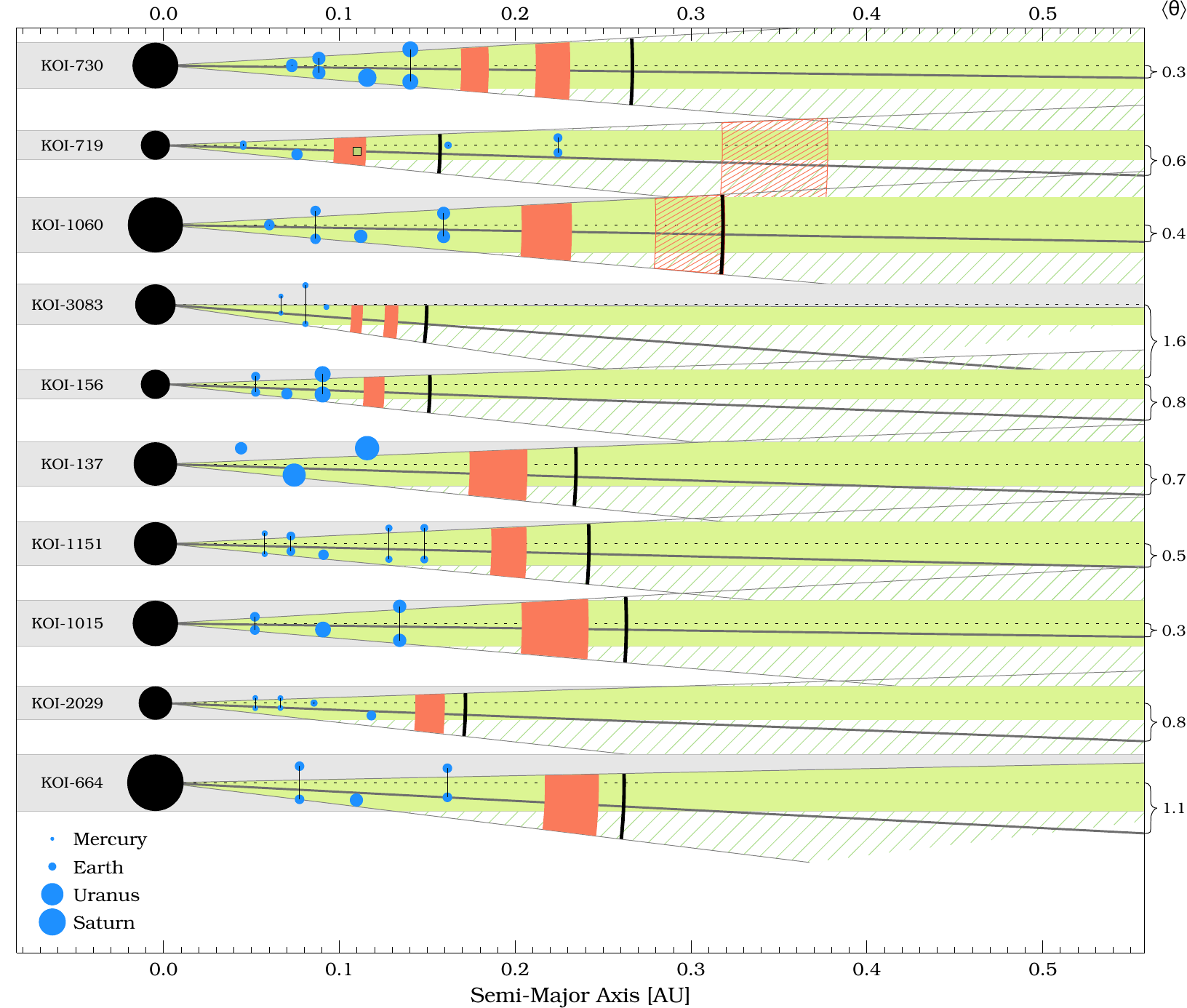}
\caption{The same as Figures~\ref{fig:planetspacing_inclination1} \& \ref{fig:planetspacing_inclination2} but for the remaining 10 systems in our sample which have at least one planet with $P_\text{trans}\ge0.55$.} 
\label{fig:planetspacing_inclination3} 
\end{figure*}

We now make updated and new TB relation predictions in all 151 systems in our sample. 
If the detected planets in a system adhere to the TB relation better than the Solar System planets ($\chi^2/\text{d.o.f.}<1.0$, Equation 4 of BL13), we only predict an 
extrapolated planet, beyond the outermost detected planet. 
If the detected planets adhere worse than the Solar System, we 
simulate the insertion of up to 9 hypothetical planets into the system, covering all possible locations and combinations, 
and calculate a new $\chi^2/\text{d.o.f.}$ value for each possibility. 
We determine how many planets to insert, and where to insert them, based on the solution which improves the system's adherence to the TB relation, 
scaled by the number of inserted planets squared. 
This protects against overfitting (inserting too many planets, resulting in too good a fit). 
In Eq. 5 of BL13 we introduced a parameter $\gamma$, which is a measure of the fractional amount by which the $\chi^2/\text{d.o.f.}$ improves, divided by the number of planets 
inserted. 
Here, we improve the definition of $\gamma$ by dividing by the square of the number of planets inserted,
\begin{equation} 
\gamma = \frac{\left(\frac{\chi_i^2-\chi_f^2}{\chi_f^2}\right)}{n_\text{ins}^2} 
\label{eq:gamma} 
\end{equation} 
where $\chi_i^2$ and $\chi_f^2$ are the $\chi^2$ of the TB relation fit before and after planets are inserted respectively, 
while $n_\text{ins}$ is the number of inserted planets.

Importantly, when we calculate our $\gamma$ value by dividing by the number of inserted planets squared, rather than the number of planets, 
we still predict the BL13 predictions that have been detected.
In two of these systems fewer planets are predicted and as a result the new 
predictions agree better with the location of the detected candidates. 
This can be seen by comparing Figures~\ref{fig:confirmed_predictions} and \ref{fig:confirmed_predictions_numins2}.

We compute $P_\text{trans}$ for each planet prediction in our sample of 151 Kepler systems.
We emphasize the 40 systems where at least one inserted planet in that system has $P_\text{trans}\ge 0.55$. 
Period predictions for this subset of 40 systems are displayed in Table~\ref{tab:predictions}1 and
Figures~\ref{fig:planetspacing_inclination1},~\ref{fig:planetspacing_inclination2} and~\ref{fig:planetspacing_inclination3}.
As discussed in the previous section, we expect a detection rate of $\sim15$ per cent for this high-$P_\text{trans}$ sample.
Predictions for all 228 planets (regardless of their $P_\text{trans}$ value) are shown in Table~\ref{tab:allpredictions}2 
(where the systems are ordered by the maximum $P_\text{trans}$ value in each system).

\subsection{Average Number of Planets in Circumstellar Habitable Zones}

\afterpage{
\begin{figure*}
\includegraphics[width=1.0\textwidth]{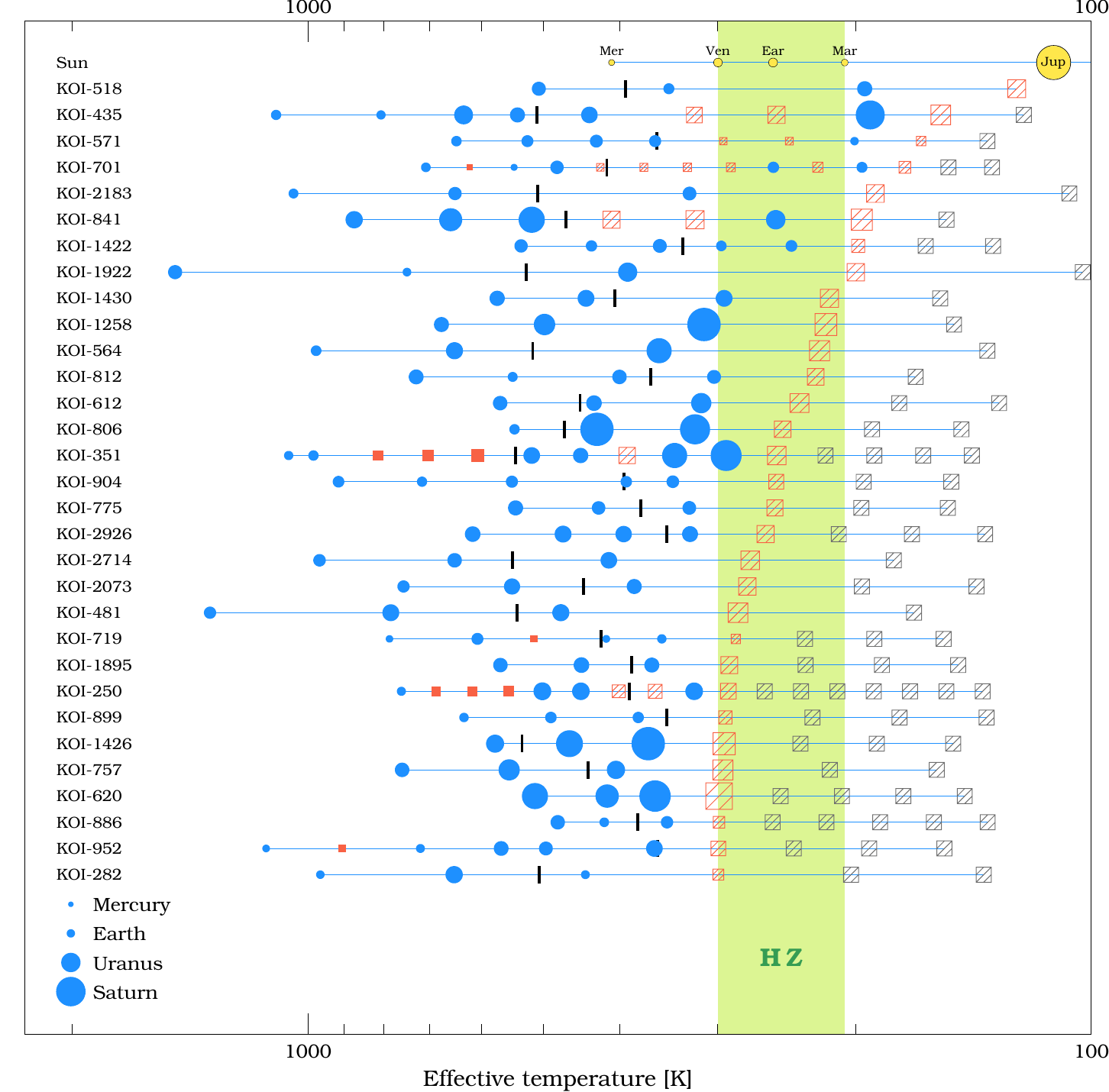}
\caption{The effective temperatures of planets within the 31 systems from our sample which extend out to the green habitable zone (HZ) 
after our planet predictions are made. 
For the purpose of estimating the number of HZ planets per star (see Table \ref{tab:numinHZ}), we extrapolate additional planets (gray squares) beyond the HZ. 
The sizes of the red hashed squares represent the $R_\text{max}$ of the predicted planet.}
\label{fig:allsystems_hz1}
\end{figure*}


\begin{table*} 
\begin{threeparttable}
\caption{\normalsize The estimated number of planets per star within various `habitable zones'. }
\renewcommand{\arraystretch}{1.2}
\setlength{\tabcolsep}{5pt}

\begin{tabular}{lcccccccc} 
 & \phantom{ab} & \multicolumn{3}{|c|}{All planets} & \phantom{abcdef} & \multicolumn{3}{|c|}{Rocky planets ($R \le 1.5\;R_\oplus$)} \\ 
\cmidrule{3-5} \cmidrule{7-9}
Sample && Mars-Venus & K13 ``optimistic'' & K13 ``conservative'' &&  Mars-Venus & K13 ``optimistic'' & K13 ``conservative'' \\ 
\hline 
All 151 systems && $2.0\pm1.0$ & $2.3\pm1.2$ & $1.5\pm0.8$ && 0.15 & 0.15 & 0.10 \\
Least extrapolation$^b$ && $1.6\pm0.9$ & $1.7\pm0.8$ & $1.3\pm0.7$ && 0.40 & 0.35 & 0.30 \\ 
\hline 
\end{tabular} 
\begin{tablenotes}
\item[a] K13 ``optimistic'' and ``conservative'' habitable zones refer to the ``recent Venus'' to ``early Mars'' and ``runaway greenhouse'' to ``maximum greenhouse' regions from \cite{Kopparapu2013} respectively'. 
\item[b] The 31 systems in the sample shown in Fig. \ref{fig:allsystems_hz1} are those which need the least extrapolation (red hashed squares)
 to extend out to (or beyond) the green Mars-Venus HZ.
\end{tablenotes}
\label{tab:numinHZ}
\end{threeparttable} 
\end{table*}
}

Since the search for earth-sized rocky planets in circumstellar habitable zones (HZ) is of particular importance, in Fig. \ref{fig:allsystems_hz1}, for a subset
of Kepler multiples whose predicted (extrapolated) planets extend to the HZ, we have converted the semi-major axes of detected and predicted planets into 
effective temperatures (as in Fig. 6 of BL13). One can see in Fig. \ref{fig:allsystems_hz1} that the habitable zone (shaded green) contains between 0 and 4 planets.
Thus, if the TB relation is approximately correct, and if Kepler multi-panet systems are representative of planetary systems in general, there are on 
average $\sim 2$ habitable zone planets per star.  

More specifically, in Table \ref{tab:numinHZ}, we estimate the number of planets per star in various `habitable zones', 
namely (1) the range of $T_\text{eff}$ between Mars and Venus (assuming an albedo of 0.3), displayed in Fig.\ref{fig:allsystems_hz1} as the green shaded region,
(2) the \cite{Kopparapu2013} ``optimistic'' and (3) ``conservative'' habitable zones 
(``recent Venus'' to ``early Mars'', and ``runaway greenhouse'' to ``maximum greenhouse'' respectively). 
We find, on average, $2 \pm 1$ planets per star in the ``habitable zone'', almost independently of which of the 3 habitable zones one is referring to.
Using our estimates of the maximum radii for these predominantly undetected (but predicted) planets, 
as well as the planetary radius distribution of Fig.~\ref{fig:planetary_radius_distribution_illustration}, 
we estimate that on average, $\sim 1/6$ of these $\sim2$ planets, or $\sim 0.3$, are `rocky'. 
We have assumed that planets with $R  \le 1.5 R_{\oplus}$ are rocky \citep{Rogers2014,Wolfgang2014}.

\section{ADJACENT PLANET PERIOD RATIOS} 
\label{sec:periodratios} 

HB14 concluded that the percentage of detected planets ($\sim 5\%$) was on the lower side of their expected range ($\sim 5\% - 20\%$) and that the TB relation may
over-predict planet pairs near the 3:2 mean-motion resonance (compared to systems which adhered to the TB relation better than the Solar System, without 
any planet insertions. i.e. $\chi^2/\text{d.o.f}\le1$). 
There is some evidence that a peak in the distribution of period ratios around the 3:2 resonance is to be expected from Kepler data, after correcting for incompleteness \cite{Steffen2014}.
In this section we investigate the period ratios of adjacent planets in our Kepler multiples before and after our new TB relation predictions are made.

We divide our sample of Kepler multiples into a number of subsets. Our first subset includes systems which adhere to the TB relation better than the Solar System 
(where we only predict an extrapolated planet beyond the outermost detected planet). 
Systems which adhere to the TB relation worse than the Solar System we divide into two subsets, before and after the planets predicted by the TB relation were inserted. 
Adjacent planet period ratios can be misleading if there is an undiscovered planet between two detected planets, which would reduce the period ratios if it was
included in the data. To minimize this incompleteness, we also construct a subset of systems which are the most likely to be completely sampled (unlikely to contain any additional transiting planets within the range of the detected planet periods).

Systems which adhere to the TB relation better than the Solar System ($\chi^2/\text{d.o.f}\le 1$) were considered by
HB14 as being the sample of planetary systems that were most complete and therefore had a distribution of 
adjacent planet period ratios most representative of actual planetary systems. 
However, the choice of BL13 to normalize the TB relation to
the Solar System's $\chi^2/\text{d.o.f}$ 
is somewhat arbitrary.
The Solar System's $\chi^2/\text{d.o.f}$ is possibly
too high to consider all those with smaller values of $\chi^2/\text{d.o.f}$ to be completely sampled.

We want to find a set of systems which are unlikely to host any additional planets between adjacent pairs, due to the system being dynamically full \citep{Hayes1998}. 
We do this by identifying the systems where two or more sequential planet pairs are likely to be unstable when a massless test particle is inserted between 
each planet pair (dynamical spacing $\Delta < 10$, \cite{Gladman1993}, BL13).

The dynamical spacing $\Delta$ is an estimate of the stability of adjacent planets. 
If inserting a test particle between a detected planet pair results in either of the two new $\Delta$ values 
being less than 10, we consider the planet pair without the insertion to be complete. 
That is, there is unlikely to be room, between the detected planet pair, where an undetected planet could exist without making the planet pair dynamically unstable.
Therefore, since the existence of an undetected planet between the planet pair is unlikely, we refer to the planet pair as `completely sampled'. 
Estimating completeness based on whether a system is dynamically full is a reasonable approach, since there is some evidence that 
the majority of systems are dynamically full (e.g. \cite{Barnes2004}). For Kepler systems in particular, \cite{Fang2013} concluded that 
{\it at least} 45 per cent of 4-planet Kepler systems are dynamically packed.

\begin{figure}
\includegraphics[width=0.47\textwidth]{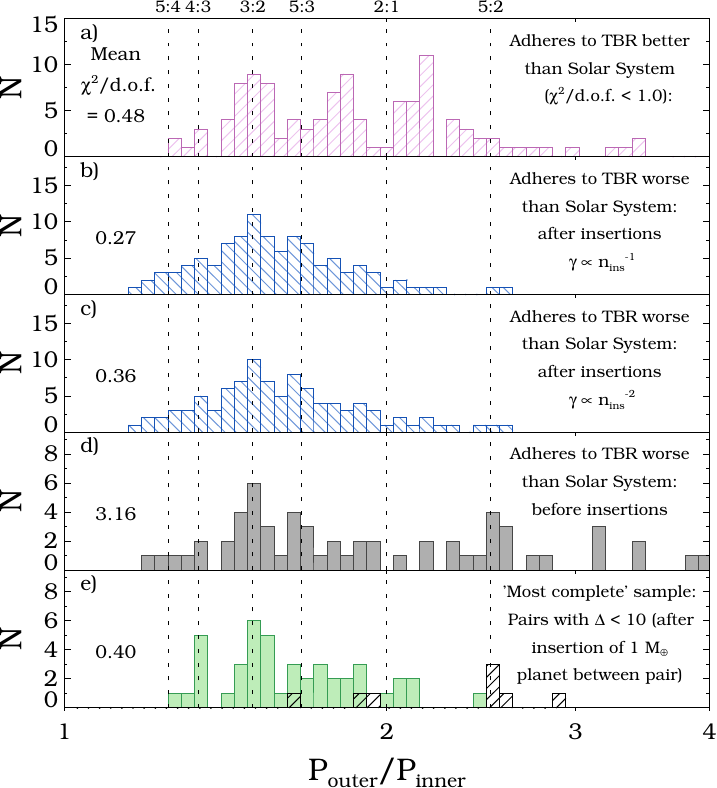}
\caption{The period ratios of adjacent-planet pairs in our sample of Kepler multiples, which can be compared to Figure 4 of HB14. 
Panel a) represents systems which adhere to the TB relation better than the Solar System $(\chi^2/\text{d.o.f.} < 1)$.
Panel b) represents those systems which adhere to the TB relation worse than the Solar System and where BL13 inserted planets.
This panel shows the period ratios of adjacent-planet pairs after planets are inserted. 
Panel c) is similar to Panel b), except that the $\gamma$ value, which is used to determine the best TB relation insertion for a given system, 
is divided by the number of inserted planets squared.
In BL13 and panel b), $\gamma$ was divided by the number of inserted planets.
Panel d) shows the period ratios between adjacent pairs of the same systems from panels b) and c), except before the additional planets from the predictions of BL13 have been inserted.
Panel e) represents our most complete sample and contains the systems which are more likely to be dynamically full 
(as defined in Section~\ref{sec:periodratios}).
The mean $\chi^2/\text{d.o.f.}$ value for each subset is shown on the left side of each panel.
Kepler's bias toward detecting compact systems dominated by short period planets may explain why the Solar System's adjacent period pairs (black hatched histogram in panel e) are not 
representative of the histogram in panel e.
The periods of predicted planets are drawn randomly from their TB relation predicted Gaussian distributions (Tables~\ref{tab:predictions}1 and~\ref{tab:predictions}2).}
\label{fig:perratio_complete_incomplete_1col}
\end{figure}

If at least two sequential adjacent-planet pairs (at least three sequential planets) satisfy this criteria, we add the subset of the system which satisfies this criteria to our `most complete'
sample. 
We use this sample to analyze the period ratios of Kepler systems.
The period ratios of the different samples described above are shown in Figure~\ref{fig:perratio_complete_incomplete_1col}.

One criticism from HB14 was that the TB relation from BL13 inserted too many planets. 
To address this criticism we have redefined $\gamma$ to be divided by the number of inserted planets squared (denominator of Eq.~\ref{eq:gamma}). 
This introduces a heavier penalty for inserting planets. 
Figure~\ref{fig:perratio_complete_incomplete_1col} displays the distributions of period ratios when using the $\gamma$ from BL13 and the new $\gamma$ of 
Equation~\ref{eq:gamma} 
(panels b and c respectively).
When using our newly defined $\gamma$, the mean $\chi^2/\text{d.o.f.}$ (displayed on the left side of the panels), more closely resembles that 
of our `most complete sample' (panel e). 

\begin{figure}
\includegraphics[width=0.47\textwidth]{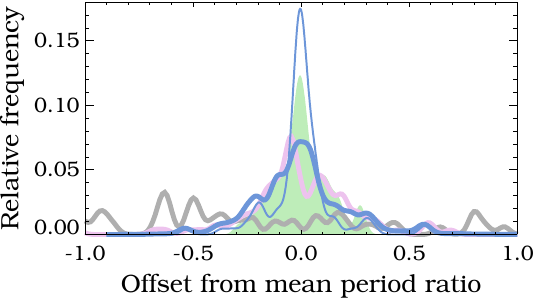}
\caption{Each system has a mean value for the adjacent-planet period ratios within that system.
This figure shows the distribution of the period ratios, offset from the mean period ratio of the system. 
The colors of the distributions correspond to subsets in Figure~\ref{fig:perratio_complete_incomplete_1col}. 
The green distribution is from the `most complete sample' in panel e) and is our best estimate of what a distribution should look like
if an appropriate number of planets have been inserted.
The gray distribution indicates that the sampling of these systems is highly incomplete. 
The thick and thin blue lines represent panel c) in two different ways.
The thick blue line uses planet periods drawn randomly from a normal distribution centered on the periods predicted by the TB relation, with the width
set to the uncertainty in the period predictions (Tables~\ref{tab:predictions}1 and~\ref{tab:predictions}2).
The thin blue line uses periods at their exact predicted value.}
\label{fig:dominant_ratio_offset}
\end{figure}

Since each panel in Fig. \ref{fig:perratio_complete_incomplete_1col} represents a mixture of planetary systems with different distributions of period ratios,
Figure \ref{fig:dominant_ratio_offset} may be a better way to compare these different samples and their adherence to the TB relation.
For each planetary system in each panel in Fig. \ref{fig:perratio_complete_incomplete_1col}, we compute the mean adjacent-planet period ratio.
Figure \ref{fig:dominant_ratio_offset} shows the distribution of the offsets from the mean period ratio of each system.
How peaked a distribution is, is a good measure of how well that distribution adheres to the TB relation.
A delta function peak at an offset of zero, would be a perfect fit. 
The period ratios of adjacent-planet pairs in our dynamically full ``most complete sample'' 
(green in Fig.  \ref{fig:dominant_ratio_offset}) displays a significant tendency to
cluster around the mean ratios. 
This clustering is the origin of the usefulness of the TB relation to predict the existence of undetected planets.
The proximity of the thick blue curve to the green distribution is a measure of how well our TB predictions can correct for
the incompleteness in Kepler multiple-planet systems and make predictions about the probable locations of the undetected planets. 

\section{CONCLUSION}
\label{sec:conclusion}

\cite{Huang2014} investigated the TB relation planet predictions of \cite{Bovaird2013} and found a detection rate of $\sim 5\%$ (5 detections from 97 predictions). 
Apart from the detections by HB14, only one additional planet (in KOI-1151) has been discovered in any of the 60  Kepler systems 
analyzed by BL13 -- indicating the advantages of such predictions while searching for new planets.
Completeness is an important issue (e.g. Figures~\ref{fig:petigura_completeness} \& \ref{fig:SNRRpcontour}).
Some large fraction of our predictions will not be detected because the planets in this fraction are likely to be too small to produce signal-to-noise ratios 
above some chosen detection threshold.
Additionally, the predicted planets may have inclinations and semi-major axes too large to transit their star as seen from Earth.
All new candidate detections based on the predictions of BL13 are approximately Earth-sized or smaller (Table \ref{tab:haungdetectiontable}). 

For a new sample of Kepler multiple-exoplanet systems containing at least three planets, 
we computed invariable plane inclinations and assumed a Gaussian opening angle of coplanarity of $\sigma_{\Delta\theta} =1.5^{\circ}$.
For each of these systems we applied an updated generalized TB relation, developed in BL13, resulting in 228 predictions in 151 systems. 

We emphasize the planet predictions which have a high geometric probability to transit, $P_\text{trans} \ge 0.55$ (Figure~\ref{fig:Ptrans_histo}). 
This subset of predictions has 77 predicted planets in 40 systems.
We expect the detection rate in this subset to be a factor of $\sim 3$ higher than the detection rate of the BL13 predictions.
 From the 40 systems with planet predictions in this sample, 24 appeared in BL13. These predictions have been updated and reprioritized. 
We have ordered our list of predicted planets based on each planet's geometric probability to transit (Tables~\ref{tab:predictions}1 and~\ref{tab:predictions}2).
Our new prioritized predictions should help on-going planet detection efforts in Kepler multi-planet systems.

\section*{Acknowledgements}
T.B. acknowledges support from an Australian Postgraduate Award.
We acknowledge useful discussions with Daniel Bayliss, Michael Ireland, George Zhou and David Nataf.

{\small
\bibliographystyle{mn2e} 
\bibliography{coplanarity_refs}

\begin{thebibliography}{}
 \providecommand{\href}[2]{#2}
  \providecommand{\doi}[1]{\href{http://dx.doi.org/#1}{doi:#1}}
  \providecommand{\eprint}[1]{\href{http://arxiv.org/abs/#1}{arXiv:#1}}

\bibitem[\protect\citeauthoryear{Ballard \& Johnson}{Ballard \&
  Johnson}{2014}]{Ballard2014}
Ballard S.,  Johnson J.~A.,  2014, ApJ (submitted), \eprint{1410.4192}

\bibitem[\protect\citeauthoryear{Barnes \& Raymond}{Barnes \&
  Raymond}{2004}]{Barnes2004}
Barnes R.,  Raymond S.~N.,  2004, ApJ, 617, 569, \doi{10.1086/423419}

\bibitem[\protect\citeauthoryear{Bovaird \& Lineweaver}{Bovaird \&
  Lineweaver}{2013}]{Bovaird2013}
Bovaird T.,  Lineweaver C.~H.,  2013, MNRAS, 435, 1126

\bibitem[\protect\citeauthoryear{Christiansen et~al.,}{Christiansen
  et~al.}{2012}]{Christiansen2012}
Christiansen J.~L.  et~al., 2012, PASP, 124, 1279

\bibitem[\protect\citeauthoryear{Dong \& Zhu}{Dong \& Zhu}{2013}]{Dong2013}
Dong S.,  Zhu Z.,  2013, ApJ, 778, 53, \doi{10.1088/0004-637X/778/1/53}

\bibitem[\protect\citeauthoryear{Dressing \& Charbonneau}{Dressing \&
  Charbonneau}{2013}]{Dressing2013}
Dressing C.~D.,  Charbonneau D.,  2013, ApJ, 767, 95,
  \doi{10.1088/0004-637X/767/1/95}

\bibitem[\protect\citeauthoryear{Fabrycky \& Winn}{Fabrycky \&
  Winn}{2009}]{Fabrycky2009}
Fabrycky D.~C.,  Winn J.~N.,  2009, ApJ, 696, 1230

\bibitem[\protect\citeauthoryear{Fabrycky et~al.,}{Fabrycky
  et~al.}{2014}]{Fabrycky2014}
Fabrycky D.~C.  et~al., 2014, ApJ, 790, 146

\bibitem[\protect\citeauthoryear{Fang \& Margot}{Fang \&
  Margot}{2012}]{Fang2012a}
Fang J.,  Margot J.-L.,  2012, ApJ, 761, 92

\bibitem[\protect\citeauthoryear{Fang \& Margot}{Fang \&
  Margot}{2013}]{Fang2013}
Fang J.,  Margot J.-L.,  2013, ApJ, 767, 115

\bibitem[\protect\citeauthoryear{Figueira et~al.,}{Figueira
  et~al.}{2012}]{Figueira2012}
Figueira P.  et~al., 2012, A\&A, 541, A139

\bibitem[\protect\citeauthoryear{Foreman-Mackey, Hogg, Morton \&
  .}{Foreman-Mackey et~al.}{2014}]{Foreman-Mackey2014}
Foreman-Mackey D.,  Hogg D.~W.,  Morton T.~D.,    . 2014, ApJ, 795, 64,
  \doi{10.1088/0004-637X/795/1/64}

\bibitem[\protect\citeauthoryear{Fressin et~al.,}{Fressin
  et~al.}{2013}]{Fressin2013}
Fressin F.  et~al., 2013, ApJ, 766, 81

\bibitem[\protect\citeauthoryear{Gladman}{Gladman}{1993}]{Gladman1993}
Gladman B.,  1993, Icarus, 106, 247

\bibitem[\protect\citeauthoryear{Hayes \& Tremaine}{Hayes \&
  Tremaine}{1998}]{Hayes1998}
Hayes W.,  Tremaine S.,  1998, Icarus, 135, 549, \doi{10.1006/icar.1998.5999}

\bibitem[\protect\citeauthoryear{Howard et~al.,}{Howard
  et~al.}{2012}]{Howard2012}
Howard A.~W.  et~al., 2012, ApJS, 201, 15

\bibitem[\protect\citeauthoryear{Huang \& Bakos}{Huang \&
  Bakos}{2014}]{Huang2014}
Huang C.~X.,  Bakos G.~A.,  2014, MNRAS, 681, 674

\bibitem[\protect\citeauthoryear{Jaki}{Jaki}{1972}]{Jaki1972}
Jaki S.~L.,  1972, Am. J. Phys., 40, 1014

\bibitem[\protect\citeauthoryear{Johansen, Davies, Church \& Holmelin}{Johansen
  et~al.}{2012}]{Johansen2012}
Johansen A.,  Davies M.~B.,  Church R.~P.,    Holmelin V.,  2012, ApJ, 758, 39,
  \doi{10.1088/0004-637X/758/1/39}

\bibitem[\protect\citeauthoryear{Kopparapu et~al.,}{Kopparapu
  et~al.}{2013}]{Kopparapu2013}
Kopparapu R.~K.  et~al., 2013, ApJ, 765, 131, \doi{10.1088/0004-637X/765/2/131}

\bibitem[\protect\citeauthoryear{Kov\'{a}cs \& Bakos}{Kov\'{a}cs \&
  Bakos}{2005}]{Kovacs2005}
Kov\'{a}cs G.,  Bakos G.,  2005, \eprint{0508081}

\bibitem[\protect\citeauthoryear{Lillo-Box, Barrado, Bouy \& .}{Lillo-Box
  et~al.}{2014}]{Lillo2014}
Lillo-Box J.,  Barrado D.,  Bouy H.,    . 2014, A\&A, 556, A103

\bibitem[\protect\citeauthoryear{Lissauer et~al.,}{Lissauer
  et~al.}{2011}]{Lissauer2011a}
Lissauer J.~J.  et~al., 2011, ApJS, 197, 8

\bibitem[\protect\citeauthoryear{Morton \& Swift}{Morton \&
  Swift}{2014}]{Morton2014}
Morton T.~D.,  Swift J.,  2014, ApJ, 791, 10, \doi{10.1088/0004-637X/791/1/10}

\bibitem[\protect\citeauthoryear{Petigura, Howard, Marcy \& .}{Petigura
  et~al.}{2013a}]{Petigura2013b}
Petigura E.~A.,  Howard A.~W.,  Marcy G.~W.,    . 2013a, PNAS, 110, 19273

\bibitem[\protect\citeauthoryear{Petigura, Marcy, Howard \& .}{Petigura
  et~al.}{2013b}]{Petigura2013}
Petigura E.~A.,  Marcy G.~W.,  Howard A.,    . 2013b, ApJ, 770, 69

\bibitem[\protect\citeauthoryear{Rogers}{Rogers}{2014}]{Rogers2014}
Rogers L.~a.,  2014, ApJ (submitted), \eprint{1407.4457}

\bibitem[\protect\citeauthoryear{Seager \& Mall\'{e}n-Ornelas}{Seager \&
  Mall\'{e}n-Ornelas}{2003}]{Seager2003}
Seager S.,  Mall\'{e}n-Ornelas G.,  2003, ApJ, 585, 1038

\bibitem[\protect\citeauthoryear{Silburt, Gaidos, Wu \& .}{Silburt
  et~al.}{2014}]{Silburt2014}
Silburt A.,  Gaidos E.,  Wu Y.,    . 2014, preprint (arXiv:1406.6048v2),
  \eprint{arXiv:1406.6048v2}

\bibitem[\protect\citeauthoryear{Souami \& Souchay}{Souami \&
  Souchay}{2012}]{Souami2012}
Souami D.,  Souchay J.,  2012, A\&A, 543, A133

\bibitem[\protect\citeauthoryear{Steffen \& Hwang}{Steffen \&
  Hwang}{2014}]{Steffen2014}
Steffen J.~H.,  Hwang J.~A.,  2014, MNRAS (submitted),
  \eprint{arXiv:1409.3320v1}

\bibitem[\protect\citeauthoryear{Tremaine \& Dong}{Tremaine \&
  Dong}{2012}]{Tremaine2012}
Tremaine S.,  Dong S.,  2012, AJ, 143, 94, \doi{10.1088/0004-6256/143/4/94}

\bibitem[\protect\citeauthoryear{Watson}{Watson}{1982}]{Watson1982}
Watson G.~S.,  1982, Journal of Applied Probability, 19, 265

\bibitem[\protect\citeauthoryear{Weissbein \& Steinberg}{Weissbein \&
  Steinberg}{2012}]{Weissbein2012}
Weissbein A.,  Steinberg E.,  2012, \eprint{arXiv:1203.6072v2}

\bibitem[\protect\citeauthoryear{Winn \& Fabrycky}{Winn \&
  Fabrycky}{2014}]{Winn2014}
Winn J.~N.,  Fabrycky D.~C.,  2014, ARAA (submitted), \eprint{1410.4199}

\bibitem[\protect\citeauthoryear{Wolfgang \& Lopez}{Wolfgang \&
  Lopez}{2014}]{Wolfgang2014}
Wolfgang A.,  Lopez E.,  2014, ApJ (submitted), \eprint{1409.2982}

\end{thebibliography}
}


\appendix

\section{Estimation of the  invariable plane of exoplanet systems}
\label{app:invplane_calculation}

\subsection{Coordinate System}
\label{app:coordinates}

In Fig.~\ref{fig:plspac_inc_single}a and this appendix we set up and explain the coordinate system used in our analysis.
The invariable plane of a planetary system can be defined as the plane passing through the barycenter of the system and is perpendicular to 
the sum $\langle \vec{\bf{L}}\rangle$, of all planets in the system:
\begin{equation} 
	\label{eq:bfL} 
	\langle \vec{\bf{L}}\rangle=\sum_j \vec{\bf{L}}_j,
\end{equation}
where $\vec {\bf{L}}_j = (L_{x},L_{y},L_{z})$ is the orbital angular momentum of the $j$th planet.
One can introduce a coordinate system in which the $x$ axis points from the system to the observer (Fig.\ref{fig:plspac_inc_single}a).
With an $x$ axis established, we are free to choose the direction of the $z$ axis. 
For example, consider the vector $\vec{\bf L}_j$ in Fig.\ref{fig:plspac_inc_single}a.
If we choose a variety of $z^{\prime}$ axes, all perpendicular to our $x$ axis, then independent of the $z^{\prime}$ axis, the quantity $\sqrt{L^{2}_{y^{\prime}} + L^{2}_{z^{\prime}}}$  is a constant.  
Thus, without loss of generality, we could choose a $z^{\prime}$ axis such that $L_{y^{\prime}} = 0$.
In Fig.\ref{fig:plspac_inc_single}a, we have choosen the $z$ axis such that the sum of the $y-$components of the angular momenta of all the planets, is zero:
\begin{equation} 
	\label{eq:L} 
	\langle \vec{L}\rangle=\sum_j \vec{L_j} = (\,\langle \vec{L} \rangle_{x},\; 0,\; \langle \vec{L} \rangle_{z}\,).
\end{equation}
In other words we have choosen the $z$ axis such that the vector defining the invariable plane, $\langle \vec{L} \rangle$, is in the $x-z$ plane.
We define the plane perpendicular to this vector as the invariable plane of the system.

The angular separation between  $\vec{\bf L}_j$ and $\vec{\langle L\rangle}$ is $\phi_j$.
$\phi_{j}$ is a positive-valued random variable and can be well-represented by a Rayleigh distribution  of mode $\sigma_\phi$ \citep{Fabrycky2009}.
For the $j$th planet, $\vec{L_j}$ is the projection of $\vec{\bf L}_j$ onto the x-z plane. 
The angle between $\vec{L_j}$ and the $z$ axis is $\theta_j$.
The angle between $\vec{\langle L\rangle}$ and the $z$ axis is $\langle \theta \rangle$ where,
\begin{equation} 
	\label{eq:theta} 
	\langle \theta \rangle = \frac{\sum_j \theta_j L_j}{\sum L_j}.
\end{equation} 
The angular separation in the $x-z$ plane between $\vec{L_j}$ and $\vec{\langle L\rangle}$ is $\Delta \theta_j$. 
In the $x-z$ plane, we then have the relation (Fig.\ref{fig:plspac_inc_single}a),
\begin{equation}
	\label{eq:deltatheta} 
	\langle\theta\rangle + \Delta\theta_j =  \theta_j ,
\end{equation} 
where $\Delta \theta_{j}$ is a normally distributed variable centered around $\langle \theta \rangle$ with a mean of $0$.
In other words, $\Delta \theta_{j}$ can be positive or negative.
A positive definite variable such as $\phi_j$ is Rayleigh distributed if it can be described as the sum of the squares of two independent normally distributed 
variables \citep{Watson1982}, i.e.  $\phi_{j} = \sqrt{\Delta \theta_{j}^{\,2} +\Delta \theta_{j,(y-z)}^{\,2}}$ where $\Delta \theta_{j,(y-z)}$ is the 
unobservable component of $\phi_{j}$ in the $y-z$ plane perpendicular to $\Delta \theta_{j}$ (see Fig.\ref{fig:plspac_inc_single}a).
From this relationship, the Gaussian distribution of $\Delta \theta_{j}$ has a standard deviation 
equal to the mode of the Rayleigh distribution of $\phi_j$: $\sigma_{\Delta \theta} = \sigma_\phi$ .

We can illustrate the meaning of the phrase ``mutual inclination'' used in the literature (e.g. \cite{Fabrycky2014}).
For example, in Fig. \ref{fig:plspac_inc_single}a, imagine adding the angular momentum vector $\vec{\bf L}_m$ of another planet. And
projecting this vector onto the $x-z$ plane and call the projection $\vec{L_m}$ (just as we projected $\vec{\bf L}_j$ into $\vec{L_j}$).
Now we can define two ``mutual inclinations'' between the orbital planes of these two planets.
$\psi_{3D}$ is the angle between $\vec{\bf L}_j$ and $\vec{\bf L}_m$ and
$\psi$ is the angle in the $x-z$ plane between $\vec{L_j}$ and $\vec{L_m}$ (i.e. $|\Delta\theta_{j}-\Delta\theta_{m}|$).

Since both $\Delta\theta_{j}$ and $\Delta\theta_{m}$ are Gaussian distributed with mean $\mu=0$, 
their difference $\Delta\theta_{j}-\Delta\theta_{m}$ is Gaussian distributed with 
$\sigma_{(\Delta\theta_{j}-\Delta\theta_{m})} = \sqrt{\sigma_{\Delta\theta_{j}}^2+\sigma_{\Delta\theta_{m}}^2}=\sqrt{2}\sigma_{\Delta\theta}$ 
and $\mu_{(\Delta\theta_{j}-\Delta\theta_{m})}=0$. 
Hence $\psi=|\Delta\theta_{j}-\Delta\theta_{m}|$ is a positive-definite half-normal Gaussian 
with mean $\mu_{\psi}= \sqrt{2/\pi} \:\; \sigma_{(\Delta\theta_{j}-\Delta\theta_{m})}=\frac{2}{\sqrt{\pi}}\sigma_{\Delta\theta}$. 

For $\psi_{3D}$, the angle between $\vec{\bf L}_j$ and $\vec{\bf L}_m$, we have,
\begin{equation}
	\label{eq:psi}
	\psi_{3D} = \sqrt{(\Delta\theta_{j}-\Delta\theta_{m})^2+(\Delta\theta_{j,(y-z)}-\Delta\theta_{m,(y-z)})^2}.
\end{equation} 
From above, $(\Delta\theta_{j}-\Delta\theta_{m})$ is Gaussian distributed with $\sigma_{(\Delta\theta_{j}-\Delta\theta_{m})}=\sqrt{2}\sigma_{\Delta\theta}$. 
Since we expect $\sigma_{(\Delta\theta_{j,(y-z)}-\Delta\theta_{m,(y-z)})} = \sigma_{(\Delta\theta_{j}-\Delta\theta_{m})}$, 
$\psi_{3D}$ is Rayleigh distributed with mode $\sigma_{i}$ (reported in Table \ref{tab:prev_studies}).
That is, $\sigma_{\psi 3D} = \sigma_{i} = \sqrt{2} \; \sigma_{\Delta\theta}$. The mean of the Rayleigh distribution of $\psi_{3D}$ is
$\mu_{\psi_{3D}}=\sqrt{\pi}\;\sigma_{\Delta\theta}$.
On average we will have $\mu_{\psi_{3D}}/\mu_{\psi}=\frac{\pi}{2}$.

%

\begin{table*}
\begin{threeparttable}
	\caption{\normalsize Comparison of exoplanet coplanarity studies}
	\label{tab:prev_studies}
	\setlength{\tabcolsep}{4pt}
	\begin{tabular}{ccccc}
		\hline
		Reference & $i$ Distribution & Observables & Mode$^a$ of Rayleigh Distributed Mutual Inclinations & Sample (quarter, multiplicity)\\
		\midrule
		\cite{Lissauer2011a} & Rayleigh & $N_p^{\,b}$ & $\sigma_i\sim2.0^\circ$ & Kepler (Q2, 1-6) \\
		\cite{Tremaine2012} & Fisher & $N_p$ & $\sigma_i^c <4.0^\circ$ & RV \& Kepler (Q2, 1-6) \\
		\cite{Figueira2012} & Rayleigh & $N_p$ & $\sigma_i^d \sim 1.4^\circ$ & HARPS \& Kepler (Q2, 1-3) \\
		\cite{Fang2012a} & Rayleigh, R of R & $N_p$, $\xi^{e}$ & $\sigma_i^c\sim1.4^\circ$ & Kepler (Q6, 1-6) \\
		\cite{Johansen2012} & uniform $i$ + rotation$^{\,f}$ & $N_p$ & $\sigma_i<3.5^\circ$ & Kepler (Q6, 1-3) \\
		\cite{Weissbein2012} & Rayleigh & $N_p$ & no fit & Kepler (Q6, 1-6) \\
		\cite{Fabrycky2014} & Rayleigh & $N_p$, $\xi$ & $\sigma_i\sim1.8^\circ$ & Kepler (Q6, 1-6) \\
		\cite{Ballard2014} & Rayleigh & $N_p$ & $\sigma_i = 2.0^{\circ\,+4.0}_{\;\;\;-2.0}$ & Kepler M-dwarfs (Q16, 2-5) \\
		\hline
	\end{tabular}
	\begin{tablenotes}[]
	\item[a] The mode $\sigma_i$ is equal to the $\sigma_{\psi_{3D}}$ discussed at the end of Appendix \ref{app:coordinates}. 
	Thus $\sigma_{i} = \sigma_{\psi 3D} = \sqrt{2} \; \sigma_{\Delta\theta}$. Assuming $\sigma_{\Delta\theta} = 1.5^{\circ}$ is equivalent to assuming $\sigma_{i}=2.1^{\circ}$.
		\item[b] $N_p$ is the multiplicity vector for the numbers of observed n-planet systems, 
		i.e. $N_p =$ (\# of 1-planet systems, \# of 2-planet systems,  \# of 3-planet systems,...).
		\item[c] Converted from the mean $\mu$ of the mutual inclination Rayleigh distribution: $\sigma_i=\sqrt{2/\pi} \:\: \mu$.
		\item[d] Converted from Rayleigh distribution relative to the invariable plane:  $\sigma_i=\sqrt{2} \;\sigma_{\Delta\theta}$.
		\item[e] $\xi$ is the normalized transit duration ratio as given in Eq. 11 of \cite{Fang2012a}.
		\item[f] Each planet is given a random uniform inclination between $0^\circ-5^\circ$. This orbital plane is then rotated uniformally between $0-2\pi$ to 
		give a random longitude of ascending node.
         \end{tablenotes}
\end{threeparttable}
\end{table*}

\subsection{Exoplanet invariable planes: permuting planet inclinations}
\label{app:permutations}

\begin{figure}
\includegraphics[width=0.47\textwidth]{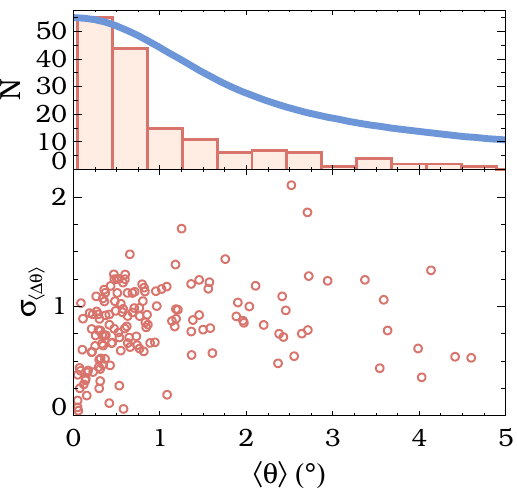}
\caption{Top panel: The red histogram represents the final $\langle\theta\rangle$ angles for the systems in our sample. 
The blue line represents the normalised area out to $a_\text{crit}$ as a function of $\langle\theta\rangle$ (Eq. \ref{eq:integral_area}). 
The bottom panel shows the dispersion for our Kepler multiples, about the calculated $\langle\theta\rangle$ value for each system.
The smaller values of $\sigma_{\langle\Delta\theta\rangle}$ at $\langle\theta\rangle \sim 0^{\circ}$ are probably due to the over-representation of
$i = 90^{\circ}$ in the Kepler data.}
\label{fig:iinvplanepdf_dispersion}
\end{figure}

An $N$-planet system has $N$ different values of $\theta_j$ (see Fig. \ref{fig:plspac_inc_single}).
Since observations are only sensitive to $|\theta_j|$, we don't know whether we are dealing with positive or negative angles. 
To model this uncertainty, we analyse the $2^{N-1}$ unique sets of permutations for positive and negative $\theta_j$ values.
For example, in a 4-planet system consider the planet with the largest angular momentum. 
We set our coordinate system by assuming its inclination $i$ is less than $90^\circ$. 
We do not know whether the $i$ values of the other 3 planets are on the same side or the opposite side of $90^{\circ}$.
The permutations of the $+1$s and $-1$s in Eq. \ref{eq:M} represent this uncertainty.
There will be $k_{max} =2^{3}=8$ sets of permutations for the $\theta_j$ of the remaining 3 planets, defined by $\theta_j M_{j,k}$ where $M_{j,k}$ is the permutation matrix,

\begin{equation}
M_{j,k} = 
 \begin{bmatrix}
1 & 1 & 1 \\
1 & 1 & -1 \\
1 & -1 & 1 \\
1 & -1 & -1 \\
-1 & 1 & 1 \\
-1 & 1 & -1 \\
-1 & -1 & 1 \\
-1 & -1 & -1 \\
 \end{bmatrix}
 .
 \label{eq:M}
\end{equation}
For each permutation $k$, we compute a notional invariant plane, by taking the angular-momentum-weighted average of the permuted angles (compare Eq. \ref{eq:theta}):

\begin{equation}
\langle\theta\rangle_{k}=\frac{\sum L_j \theta_{j}M_{j,k}}{\sum L_j}
\label{eq:invplane_median_app}
\end{equation}
which yields 8 unique values of $\langle\theta\rangle_{k}$, each consistent with the $b_j$, $\theta_j$ and $i_j$ values from the transit light curves (see Eqs.\ref{eq:b},\ref{eq:b2},\ref{eq:i}).
For each of these $\langle\theta\rangle_{k}$, we compute a proxy for coplanarity which is the mean of the angular-momentum-weighted angle of the orbital planes around the notional invariant plane: 
\begin{equation}
\sigma_{\langle\Delta \theta \rangle_k} = \frac{\sum L_j |\langle\theta\rangle_k - \theta_j M_{j,k} |}{\sum L_j}. 
\label{eq:deltathetak}
\end{equation}
The smaller the value of $\sigma_{\langle\Delta \theta \rangle_k}$, the more coplanar that permutation set is.
This permutation procedure is most appropriate when the system is close to edge-on since in this case the various planets are equally likely to have actual inclinations 
on either side of $90^{\circ}$.
By contrast, when $\langle \theta \rangle$ is large, these permutations exaggerate the uncertainty since most planets are likely to be on the same side of 
$90^{\circ}$ as the dominant planet. 
Thus, using this method, close-to-edge-on systems with $\langle \theta \rangle \lesssim 0.5^{\circ}$ will yield the smallest and more appropriate dispersions which we find to be in the 
range:  $0^{\circ} \lesssim \sigma_{\langle\Delta \theta \rangle_k} \lesssim 1.5^{\circ} $.
Since the coplanarity of a system should not depend on the angle to the observer, the values of $\sigma_{\langle \Delta \theta \rangle_k}$ should not 
depend on $\langle\theta\rangle$. We find that this condition can best be met when we reject permutations which yield values of $\sigma_{\langle\Delta \theta \rangle_k}$ less 
than $0.4^\circ$ and greater than $1.5^\circ$.
(see Fig.~\ref{fig:iinvplanepdf_dispersion}). 
When no permutations for a given system meet this criteria, we select the single permutation which is closest to this range.
Since the sign of $\langle\theta\rangle$ is not important, 
when more than one permutation meets this criterion, we estimate $\langle\theta\rangle$
by taking the median of the absolute values of the $\langle\theta\rangle_{k}$ 
for which $0.4^{\circ} \lesssim \sigma_{\langle\Delta \theta \rangle_k}  \lesssim 1.5^{\circ}$.
These permutations are used in Figs. \ref{fig:planetspacing_inclination1},\ref{fig:planetspacing_inclination2} \& \ref{fig:planetspacing_inclination3},
where the most probable inclination ambiguities are indicated by two blue planets at the same semi-major axis, one above and one below
the $i=90$ dashed horizontal line.

\section{Calculating the Geometric Probability to Transit: $\boldsymbol {P_\text{trans}}$}
\label{app:Ptrans}

\begin{figure}
\includegraphics[width=0.47\textwidth]{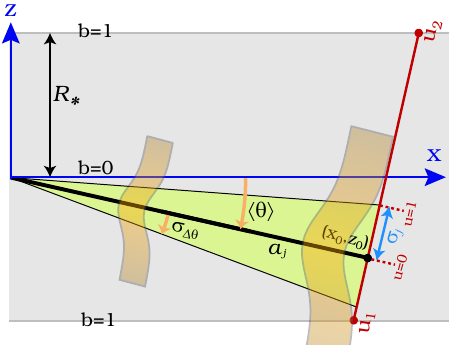}
\caption{Illustration of the variables required to calculate $P_\text{trans}$ (Eq. \ref{eq:ptrans}), 
the fraction of a Gaussian within the transit region ($b\lesssim1$).
We assume that a planet with semi-major axis $a_j$, in a system whose invariable plane is the thick line, will have a position $\Delta \theta_j$, whose probability will be 
described by a normal distribution along the red line perpendicular to the invariant plane. 
The integration variable $u$ of Eq. \ref{eq:ptrans} is along this perpendicular red line.
The mean of the Gaussian is at the point $(x_o, z_o)$ where $u = 0$.  
The standard deviation  $\sigma_j = a_j \sigma_{\Delta \theta}$ is also the half-width of the green region at a distance $a_j$ from the host star.}
\label{fig:var_illustration}
\end{figure}

We assume $\sigma_{\Delta \theta} = 1.5^{\circ}$ is constant over all systems (Section \ref{sec:exoplanet_coplanarity}), such that the geometric probability for the $j$th planet to transit 
is the fraction of a Gaussian-weighted opening angle within the transit region, where the standard deviation of the Gaussian is $\sigma_j = a_j \sigma_{\Delta \theta}$ 
(Figure~\ref{fig:var_illustration} and Eq. \ref{eq:ptrans}).
 
\begin{equation}
\label{eq:ptrans}
P_\text{trans}(\langle\theta\rangle,a_j)=\frac{1}{\sigma_j \sqrt{2\pi}}\int^{u_2}_{u_1} e^{-\tfrac{u^2}{2\sigma_{j}^{2}}}du
\end{equation}
where  the integration limits $u_1$ and $u_2$ are defined by (see Fig. \ref{fig:var_illustration}),

\begin{align}
u_1& = \frac{R_* + z_o}{\cos\langle\theta\rangle}\\
u_2& = \frac{R_* - z_o}{\cos\langle\theta\rangle}\\
z_0& = - a_{j}\sin\langle\theta\rangle.
\end{align}
with the axes in Fig.~\ref{fig:var_illustration} being the same as in panel b of Fig.~\ref{fig:plspac_inc_single}, we have $z_o < 0$.
And since $\langle\theta\rangle \sim 0$, we have $\cos\langle\theta\rangle \sim 1$.

In the panels of Figs. \ref{fig:planetspacing_inclination1}, \ref{fig:planetspacing_inclination2} and \ref{fig:planetspacing_inclination3},
the green area is a function of the $\langle \theta \rangle$ of the system.
We can integrate $P_\text{trans}$ to get the size of the green area that is closer to the host star than $a_{crit}$. 
This yields an area that is an estimate of the amount of parameter space in which planets can transit:
\begin{equation}
\label{eq:integral_area}
Area(\langle\theta\rangle) = \int_{0}^{a_{crit}(\langle\theta\rangle)}  P_\text{trans} \: da.
\end{equation}
In the top panel of Fig. \ref{fig:iinvplanepdf_dispersion} the blue curve is a normalized version of Eq. \ref{eq:integral_area} and represents the statistical expectation of
the number of systems as a function of $\langle\theta\rangle$ (ignoring detection biases). The histogram shows the values we have obtained.

\onecolumn
\section{Tables of planet predictions}
\begin{ThreePartTable}
\footnotesize
\renewcommand{\arraystretch}{0.6}
\setlength{\tabcolsep}{8pt}
\begin{TableNotes}
\item[a] $\Delta\gamma=(\gamma_1-\gamma_2)/\gamma_2$ where $\gamma_1$ and $\gamma_2$ are the highest and second highest $\gamma$ values for that system respectively (see Bovaird \& Lineweaver (2013)).
\item[b] $R_\text{max}$ is calculated by applying the lowest SNR of the detected planets in the system to the period of the inserted planet. See Eq. \ref{eq:Rmax}.
\item[c] A planet number followed by ``E" indicates the planet is extrapolated (has a larger period than the outermost detected planet in the system).
\end{TableNotes}

\begin{longtable}{lccccccccccc}
\caption{77 Planet predictions with a high geometric probability to transit ($P_\text{trans}\ge0.55$) in 40 Kepler systems}\\

\toprule
System & Number & $\gamma$ & $\Delta\gamma^a$ & $\left(\dfrac{\chi^2}{\text{d.o.f.}}\right)_i$ & $\left(\dfrac{\chi^2}{\text{d.o.f.}}\right)_f$ & Inserted & Period & a      & R$_\text{max}$$^{\,b}$   & T$_\text{eff}$ & P$_\text{trans}$\\
 & Inserted & & & & & Planet \# & (days) & (AU) & (R$\boldsymbol{_\oplus}$) & (K) & \\
\midrule
\endhead

\cmidrule{3-3}
\multicolumn{3}{r}{\textit{continued}}
\endfoot

\bottomrule
\insertTableNotes
\endlastfoot

KOI-1198 & 2 & 2.2 & 0.3 & 7.19 & 0.74 & 1 & $2.1\pm0.4$ & 0.03 & 1.3 & 1642 & 1.00 \\
 &  &  &  &  &  & 2 & $4.3\pm0.7$ & 0.06 & 1.5 & 1297 & 1.00 \\
KOI-1955 & 4 & 3.4 & 0.2 & 10.23 & 0.19 & 1 & $2.6\pm0.3$ & 0.04 & 0.9 & 1568 & 1.00 \\
 &  &  &  &  &  & 2 & $4.1\pm0.5$ & 0.05 & 1.0 & 1347 & 1.00 \\
 &  &  &  &  &  & 3 & $6.4\pm0.7$ & 0.07 & 1.2 & 1157 & 0.99 \\
 &  &  &  &  &  & 4 & $10.1\pm1.1$ & 0.10 & 1.3 & 994 & 0.94 \\
KOI-1082 & 2 & 23.0 & 5.0 & 5.02 & 0.06 & 1 & $1.8\pm0.2$ & 0.03 & 1.0 & 1184 & 1.00 \\
 &  &  &  &  &  & 2 & $2.8\pm0.3$ & 0.04 & 1.1 & 1029 & 1.00 \\
 &  &  &  &  &  & \quad\;\;\,3 E$\,^\text{c}$ & $14.7\pm1.4$ & 0.11 & 1.6 & 588 & 0.72 \\
KOI-952 & 1 & 2.2 & 5.2 & 2.36 & 0.76 & 1 & $1.5\pm0.3$ & 0.02 & 0.9 & 904 & 1.00 \\
KOI-500 & 2 & 7.6 & 1.9 & 5.47 & 0.18 & 1 & $1.5\pm0.2$ & 0.02 & 1.2 & 1091 & 1.00 \\
 &  &  &  &  &  & 2 & $2.1\pm0.2$ & 0.03 & 1.3 & 960 & 1.00 \\
KOI-4032 & 2 & 0.9 & 1.9 & 1.20 & 0.27 & 1 & $3.4\pm0.2$ & 0.04 & 0.8 & 1347 & 1.00 \\
 &  &  &  &  &  & 2 & $4.5\pm0.2$ & 0.05 & 0.9 & 1224 & 0.99 \\
 &  &  &  &  &  & \quad3 E & $6.9\pm0.3$ & 0.07 & 1.0 & 1061 & 0.91 \\
KOI-707 & 1 & 3.2 & 2.2 & 3.69 & 0.90 & 1 & $8.9\pm0.8$ & 0.09 & 2.0 & 1162 & 1.00 \\
KOI-1336 & 2 & 1.3 & 65.6 & 1.07 & 0.19 & 1 & $6.7\pm0.7$ & 0.07 & 1.7 & 1053 & 0.98 \\
 &  &  &  &  &  & 2 & $25.1\pm2.5$ & 0.17 & 2.4 & 679 & 0.64 \\
KOI-2859 & 1 & 10.2 & -1.0 & 1.69 & 0.16 & 1 & $2.4\pm0.1$ & 0.03 & 0.6 & 1242 & 0.98 \\
KOI-250 & 5 & 0.7 & 1.1 & 2.26 & 0.14 & 1 & $4.8\pm0.4$ & 0.05 & 1.2 & 686 & 0.96 \\
 &  &  &  &  &  & 2 & $6.6\pm0.6$ & 0.06 & 1.3 & 616 & 0.91 \\
 &  &  &  &  &  & 3 & $9.2\pm0.8$ & 0.07 & 1.4 & 553 & 0.83 \\
KOI-168 & 0 & - & - & 0.14 & - & \quad 1 E & $22.6\pm1.2$ & 0.17 & 2.6 & 909 & 0.95 \\
 &  &  &  &  &  & \quad 2 E & $33.2\pm1.4$ & 0.21 & 2.9 & 800 & 0.87 \\
 &  &  &  &  &  & \quad 3 E & $48.6\pm1.7$ & 0.28 & 3.2 & 704 & 0.76 \\
 &  &  &  &  &  & \quad 4 E & $71.3\pm2.2$ & 0.36 & 3.5 & 620 & 0.64 \\
KOI-2585 & 0 & - & - & 0.55 & - & \quad 1 E & $14.8\pm0.7$ & 0.12 & 1.1 & 967 & 0.95 \\
 &  &  &  &  &  & \quad 2 E & $20.7\pm0.8$ & 0.15 & 1.2 & 865 & 0.87 \\
 &  &  &  &  &  & \quad 3 E & $28.9\pm0.9$ & 0.19 & 1.3 & 774 & 0.78 \\
 &  &  &  &  &  & \quad 4 E & $40.4\pm1.1$ & 0.24 & 1.4 & 692 & 0.67 \\
 &  &  &  &  &  & \quad 5 E & $56.4\pm1.3$ & 0.30 & 1.6 & 619 & 0.57 \\
KOI-1052 & 2 & 115.4 & 40.4 & 1.36 & 0.01 & 1 & $10.8\pm1.2$ & 0.10 & 1.6 & 909 & 0.94 \\
 &  &  &  &  &  & 2 & $27.1\pm2.8$ & 0.18 & 2.0 & 669 & 0.69 \\
KOI-505 & 3 & 1.6 & 0.1 & 8.20 & 0.56 & 1 & $21.9\pm2.5$ & 0.15 & 4.5 & 788 & 0.92 \\
 &  &  &  &  &  & 2 & $34.7\pm3.9$ & 0.20 & 5.1 & 676 & 0.78 \\
 &  &  &  &  &  & 3 & $55\pm7$ & 0.27 & 5.7 & 580 & 0.62 \\
KOI-1831 & 1 & 1.3 & 0.2 & 2.54 & 1.11 & 1 & $8.2\pm1.1$ & 0.08 & 0.9 & 739 & 0.91 \\
KOI-248 & 1 & 18.3 & 80.8 & 2.48 & 0.13 & 1 & $4.2\pm0.5$ & 0.04 & 1.4 & 633 & 0.89 \\
KOI-880 & 1 & 4.0 & 22.4 & 1.35 & 0.28 & 1 & $11.8\pm2.0$ & 0.10 & 2.1 & 761 & 0.89 \\
KOI-1567 & 2 & 5.9 & 44.8 & 1.51 & 0.07 & 1 & $11.4\pm1.2$ & 0.10 & 2.0 & 668 & 0.89 \\
 &  &  &  &  &  & 2 & $28.0\pm2.9$ & 0.18 & 2.5 & 494 & 0.62 \\
KOI-1952 & 2 & 85.6 & 16.6 & 3.26 & 0.01 & 1 & $12.1\pm1.2$ & 0.10 & 1.5 & 828 & 0.87 \\
 &  &  &  &  &  & 2 & $18.3\pm1.8$ & 0.14 & 1.6 & 720 & 0.75 \\
KOI-351 & 4 & 0.5 & 1.0 & 5.78 & 0.65 & 1 & $15.4\pm1.7$ & 0.13 & 1.4 & 813 & 0.87 \\
 &  &  &  &  &  & 2 & $23.9\pm2.5$ & 0.17 & 1.6 & 702 & 0.74 \\
 &  &  &  &  &  & 3 & $37.1\pm3.9$ & 0.23 & 1.8 & 607 & 0.60 \\
KOI-701 & 6 & 18.0 & 7.4 & 4.04 & 0.01 & 1 & $8.4\pm0.8$ & 0.07 & 0.6 & 621 & 0.87 \\
KOI-1306 & 1 & 5.7 & 0.4 & 4.12 & 0.62 & 1 & $13.7\pm2.1$ & 0.11 & 1.5 & 756 & 0.85 \\
KOI-2722 & 0 & - & - & 0.54 & - & \quad 1 E & $23.4\pm1.5$ & 0.17 & 1.4 & 774 & 0.78 \\
 &  &  &  &  &  & \quad 2 E & $33.0\pm1.8$ & 0.21 & 1.5 & 690 & 0.67 \\
 &  &  &  &  &  & \quad 3 E & $46.5\pm2.3$ & 0.27 & 1.6 & 615 & 0.56 \\
KOI-1358 & 0 & - & - & 0.01 & - & \quad 1 E & $13.6\pm1.0$ & 0.10 & 1.6 & 522 & 0.74 \\
 &  &  &  &  &  & \quad 2 E & $21.0\pm1.3$ & 0.14 & 1.8 & 451 & 0.60 \\
KOI-1627 & 0 & - & - & 0.24 & - & \quad 1 E & $16.6\pm1.2$ & 0.13 & 1.9 & 586 & 0.73 \\
 &  &  &  &  &  & \quad 2 E & $27.4\pm1.5$ & 0.17 & 2.1 & 497 & 0.57 \\
KOI-1833 & 0 & - & - & 0.74 & - & \quad 1 E & $11.3\pm0.6$ & 0.08 & 2.0 & 514 & 0.72 \\
 &  &  &  &  &  & \quad 2 E & $16.4\pm0.7$ & 0.10 & 2.2 & 455 & 0.60 \\
KOI-3158 & 0 & - & - & 0.23 & - & \quad 1 E & $12.7\pm0.6$ & 0.09 & 0.4 & 566 & 0.71 \\
 &  &  &  &  &  & \quad 2 E & $16.4\pm0.7$ & 0.11 & 0.4 & 520 & 0.62 \\
KOI-2055 & 0 & - & - & 0.29 & - & \quad 1 E & $13.6\pm1.0$ & 0.11 & 1.3 & 703 & 0.70 \\
 &  &  &  &  &  & \quad 2 E & $20.6\pm1.2$ & 0.14 & 1.5 & 612 & 0.56 \\
KOI-245 & 3 & 1.0 & 1.3 & 1.56 & 0.17 & 1 & $16.8\pm0.9$ & 0.12 & 0.3 & 582 & 0.70 \\
 &  &  &  &  &  & 2 & $26.1\pm1.4$ & 0.16 & 0.3 & 502 & 0.56 \\
KOI-749 & 0 & - & - & 0.49 & - & \quad 1 E & $11.4\pm0.6$ & 0.10 & 1.5 & 711 & 0.69 \\
 &  &  &  &  &  & \quad 2 E & $16.4\pm0.7$ & 0.12 & 1.7 & 630 & 0.57 \\
KOI-730 & 0 & - & - & 0.33 & - & \quad 1 E & $27.9\pm1.5$ & 0.18 & 2.3 & 620 & 0.69 \\
 &  &  &  &  &  & \quad 2 E & $39.0\pm1.8$ & 0.22 & 2.5 & 554 & 0.58 \\
KOI-719 & 1 & 1.1 & 1.5 & 1.36 & 0.66 & 1 & $14.8\pm2.0$ & 0.11 & 0.8 & 514 & 0.69 \\
KOI-1060 & 0 & - & - & 0.22 & - & \quad 1 E & $32.7\pm2.6$ & 0.22 & 2.1 & 703 & 0.68 \\
KOI-3083 & 0 & - & - & 0.56 & - & \quad 1 E & $13.2\pm0.5$ & 0.11 & 0.7 & 751 & 0.66 \\
 &  &  &  &  &  & \quad 2 E & $16.9\pm0.5$ & 0.13 & 0.7 & 692 & 0.57 \\
KOI-156 & 0 & - & - & 0.10 & - & \quad 1 E & $17.9\pm1.0$ & 0.12 & 1.6 & 476 & 0.61 \\
KOI-137 & 0 & - & - & 0.13 & - & \quad 1 E & $31.2\pm3.1$ & 0.19 & 3.1 & 539 & 0.60 \\
KOI-1151 & 0 & - & - & 0.85 & - & \quad 1 E & $33.0\pm2.2$ & 0.20 & 1.0 & 564 & 0.59 \\
KOI-1015 & 0 & - & - & 0.62 & - & \quad 1 E & $36.1\pm3.5$ & 0.22 & 2.3 & 590 & 0.58 \\
KOI-2029 & 0 & - & - & 0.31 & - & \quad 1 E & $23.7\pm1.6$ & 0.15 & 0.9 & 514 & 0.56 \\
KOI-664 & 0 & - & - & 0.06 & - & \quad 1 E & $40.3\pm3.0$ & 0.23 & 1.5 & 618 & 0.56 \\

\end{longtable}
\label{tab:predictions}
\end{ThreePartTable}
\twocolumn


\onecolumn
\begin{ThreePartTable}
\footnotesize
\renewcommand{\arraystretch}{0.6}
\setlength{\tabcolsep}{8pt}
\begin{TableNotes}
\item[a] $\Delta\gamma=(\gamma_1-\gamma_2)/\gamma_2$ where $\gamma_1$ and $\gamma_2$ are the highest and second highest $\gamma$ values for that system 
respectively (see BL13).
\item[b] $R_\text{max}$ is calculated by applying the lowest SNR of the detected planets in the system to the period of the inserted planet. See Eq. \ref{eq:Rmax}.
\item[c] A planet number followed by ``E" indicates the planet is extrapolated (has a larger period than the outermost detected planet in the system).
\item[d] $P_\text{trans}$ values $\ge$ 0.55 are shown in bold, indicating a higher probability to transit.
\item[e] $T_\text{eff}$ values between Mars and Venus (206 K to 300 K, assuming an albedo of 0.3) are shown in bold.

\end{TableNotes}

\begin{longtable}{lccccccccccc}
\caption{All 228 Planet Predictions in 151 Systems (Table~\ref{tab:predictions}1 is a high $P_{\text{trans}}$ subset of this table)}\\

\toprule
System & Number & $\gamma$ & $\Delta\gamma^a$ & $\left(\dfrac{\chi^2}{\text{d.o.f.}}\right)_i$ & $\left(\dfrac{\chi^2}{\text{d.o.f.}}\right)_f$ & Inserted & Period & a  & R$_\text{max}$$^{\,b}$   & T$_\text{eff}$$^{\,e}$ & P$_\text{trans}{}^d$\\
 & Inserted & & & & & Planet \# & (days) & (AU) & (R$\boldsymbol{_\oplus}$) & (K) & \\
\midrule
\endhead

\cmidrule{3-3}
\multicolumn{3}{r}{\textit{continued}}
\endfoot

\bottomrule
\insertTableNotes
\endlastfoot

KOI-1198 & 2 & 2.2 & 0.3 & 7.19 & 0.74 & 1 & $2.1\pm0.4$ & 0.03 & 1.3 & 1642 & \bf{1.00} \\
 &  &  &  &  &  & 2 & $4.3\pm0.7$ & 0.06 & 1.5 & 1297 & \bf{1.00} \\
 &  &  &  &  &  & \quad\;\;\,3 E$\,^\text{c}$ & $73\pm12$ & 0.37 & 3.1 & 505 & 0.41 \\
KOI-1955 & 4 & 3.4 & 0.2 & 10.23 & 0.19 & 1 & $2.6\pm0.3$ & 0.04 & 0.9 & 1568 & \bf{1.00} \\
 &  &  &  &  &  & 2 & $4.1\pm0.5$ & 0.05 & 1.0 & 1347 & \bf{1.00} \\
 &  &  &  &  &  & 3 & $6.4\pm0.7$ & 0.07 & 1.2 & 1157 & \bf{0.99} \\
 &  &  &  &  &  & 4 & $10.1\pm1.1$ & 0.10 & 1.3 & 994 & \bf{0.94} \\
 &  &  &  &  &  & \quad5 E & $62\pm7$ & 0.33 & 2.0 & 541 & 0.42 \\
KOI-1082 & 2 & 23.0 & 5.0 & 5.02 & 0.06 & 1 & $1.8\pm0.2$ & 0.03 & 1.0 & 1184 & \bf{1.00} \\
 &  &  &  &  &  & 2 & $2.8\pm0.3$ & 0.04 & 1.1 & 1029 & \bf{1.00} \\
 &  &  &  &  &  & \quad3 E & $14.7\pm1.4$ & 0.11 & 1.6 & 588 & 0.72 \\
KOI-952 & 1 & 2.2 & 5.2 & 2.36 & 0.76 & 1 & $1.5\pm0.3$ & 0.02 & 0.9 & 904 & \bf{1.00} \\
 &  &  &  &  &  & \quad2 E & $40.0\pm6.2$ & 0.19 & 2.1 & \bf{299} & 0.36 \\
KOI-500 & 2 & 7.6 & 1.9 & 5.47 & 0.18 & 1 & $1.5\pm0.2$ & 0.02 & 1.2 & 1091 & \bf{1.00} \\
 &  &  &  &  &  & 2 & $2.1\pm0.2$ & 0.03 & 1.3 & 960 & \bf{1.00} \\
 &  &  &  &  &  & \quad3 E & $14.5\pm1.3$ & 0.10 & 2.2 & 506 & 0.50 \\
KOI-4032 & 2 & 0.9 & 1.9 & 1.20 & 0.27 & 1 & $3.4\pm0.2$ & 0.04 & 0.8 & 1347 & \bf{1.00} \\
 &  &  &  &  &  & 2 & $4.5\pm0.2$ & 0.05 & 0.9 & 1224 & \bf{0.99} \\
 &  &  &  &  &  & \quad3 E & $6.9\pm0.3$ & 0.07 & 1.0 & 1061 & 0.91 \\
KOI-707 & 1 & 3.2 & 2.2 & 3.69 & 0.90 & 1 & $8.9\pm0.8$ & 0.09 & 2.0 & 1162 & \bf{1.00} \\
 &  &  &  &  &  & \quad2 E & $68\pm7$ & 0.35 & 3.3 & 590 & 0.50 \\
KOI-1336 & 2 & 1.3 & 65.6 & 1.07 & 0.19 & 1 & $6.7\pm0.7$ & 0.07 & 1.7 & 1053 & \bf{0.98} \\
 &  &  &  &  &  & 2 & $25.1\pm2.5$ & 0.17 & 2.4 & 679 & \bf{0.64} \\
 &  &  &  &  &  & \quad3 E & $60\pm6$ & 0.31 & 3.0 & 507 & 0.39 \\
KOI-2859 & 1 & 10.2 & -1.0 & 1.69 & 0.16 & 1 & $2.4\pm0.1$ & 0.03 & 0.6 & 1242 & \bf{0.98} \\
 &  &  &  &  &  & \quad2 E & $5.1\pm0.3$ & 0.05 & 0.8 & 967 & 0.54 \\
KOI-250 & 5 & 0.7 & 1.1 & 2.26 & 0.14 & 1 & $4.8\pm0.4$ & 0.05 & 1.2 & 686 & \bf{0.96} \\
 &  &  &  &  &  & 2 & $6.6\pm0.6$ & 0.06 & 1.3 & 616 & \bf{0.91} \\
 &  &  &  &  &  & 3 & $9.2\pm0.8$ & 0.07 & 1.4 & 553 & \bf{0.83} \\
 &  &  &  &  &  & 4 & $24.1\pm1.9$ & 0.14 & 1.8 & 401 & 0.53 \\
 &  &  &  &  &  & 5 & $33.2\pm2.6$ & 0.17 & 2.0 & 360 & 0.44 \\
 &  &  &  &  &  & \quad6 E & $63.3\pm5.0$ & 0.27 & 2.3 & \bf{290} & 0.29 \\
KOI-168 & 0 & - & - & 0.14 & - & \quad 1 E & $22.6\pm1.2$ & 0.17 & 2.6 & 909 & \bf{0.95} \\
 &  &  &  &  &  & \quad 2 E & $33.2\pm1.4$ & 0.21 & 2.9 & 800 & \bf{0.87} \\
 &  &  &  &  &  & \quad 3 E & $48.6\pm1.7$ & 0.28 & 3.2 & 704 & \bf{0.76} \\
 &  &  &  &  &  & \quad 4 E & $71.3\pm2.2$ & 0.36 & 3.5 & 620 & \bf{0.64} \\
 &  &  &  &  &  & \quad 5 E & $104.5\pm2.7$ & 0.46 & 3.8 & 546 & 0.52 \\
KOI-2585 & 0 & - & - & 0.55 & - & \quad 1 E & $14.8\pm0.7$ & 0.12 & 1.1 & 967 & \bf{0.95} \\
 &  &  &  &  &  & \quad 2 E & $20.7\pm0.8$ & 0.15 & 1.2 & 865 & \bf{0.87} \\
 &  &  &  &  &  & \quad 3 E & $28.9\pm0.9$ & 0.19 & 1.3 & 774 & \bf{0.78} \\
 &  &  &  &  &  & \quad 4 E & $40.4\pm1.1$ & 0.24 & 1.4 & 692 & \bf{0.67} \\
 &  &  &  &  &  & \quad 5 E & $56.4\pm1.3$ & 0.30 & 1.6 & 619 & \bf{0.57} \\
KOI-1052 & 2 & 115.4 & 40.4 & 1.36 & 0.01 & 1 & $10.8\pm1.2$ & 0.10 & 1.6 & 909 & \bf{0.94} \\
 &  &  &  &  &  & 2 & $27.1\pm2.8$ & 0.18 & 2.0 & 669 & \bf{0.69} \\
 &  &  &  &  &  & \quad3 E & $110\pm20$ & 0.46 & 2.8 & 423 & 0.31 \\
KOI-505 & 3 & 1.6 & 0.1 & 8.20 & 0.56 & 1 & $21.9\pm2.5$ & 0.15 & 4.5 & 788 & \bf{0.92} \\
 &  &  &  &  &  & 2 & $34.7\pm3.9$ & 0.20 & 5.1 & 676 & \bf{0.78} \\
 &  &  &  &  &  & 3 & $55\pm7$ & 0.27 & 5.7 & 580 & \bf{0.62} \\
 &  &  &  &  &  & \quad4 E & $140\pm20$ & 0.51 & 7.2 & 426 & 0.36 \\
KOI-1831 & 1 & 1.3 & 0.2 & 2.54 & 1.11 & 1 & $8.2\pm1.1$ & 0.08 & 0.9 & 739 & \bf{0.91} \\
 &  &  &  &  &  & \quad2 E & $100\pm20$ & 0.41 & 1.7 & 316 & 0.25 \\
KOI-248 & 1 & 18.3 & 80.8 & 2.48 & 0.13 & 1 & $4.2\pm0.5$ & 0.04 & 1.4 & 633 & \bf{0.89} \\
 &  &  &  &  &  & \quad2 E & $30.1\pm3.3$ & 0.16 & 2.2 & 329 & 0.30 \\
KOI-880 & 1 & 4.0 & 22.4 & 1.35 & 0.28 & 1 & $11.8\pm2.0$ & 0.10 & 2.1 & 761 & \bf{0.89} \\
 &  &  &  &  &  & \quad2 E & $120\pm20$ & 0.45 & 3.7 & 354 & 0.27 \\
KOI-1567 & 2 & 5.9 & 44.8 & 1.51 & 0.07 & 1 & $11.4\pm1.2$ & 0.10 & 2.0 & 668 & \bf{0.89} \\
 &  &  &  &  &  & 2 & $28.0\pm2.9$ & 0.18 & 2.5 & 494 & \bf{0.62} \\
 &  &  &  &  &  & \quad3 E & $69\pm8$ & 0.32 & 3.1 & 366 & 0.37 \\
KOI-1952 & 2 & 85.6 & 16.6 & 3.26 & 0.01 & 1 & $12.1\pm1.2$ & 0.10 & 1.5 & 828 & \bf{0.87} \\
 &  &  &  &  &  & 2 & $18.3\pm1.8$ & 0.14 & 1.6 & 720 & \bf{0.75} \\
 &  &  &  &  &  & \quad3 E & $64\pm7$ & 0.31 & 2.2 & 474 & 0.38 \\
KOI-351 & 4 & 0.5 & 1.0 & 5.78 & 0.65 & 1 & $15.4\pm1.7$ & 0.13 & 1.4 & 813 & \bf{0.87} \\
 &  &  &  &  &  & 2 & $23.9\pm2.5$ & 0.17 & 1.6 & 702 & \bf{0.74} \\
 &  &  &  &  &  & 3 & $37.1\pm3.9$ & 0.23 & 1.8 & 607 & \bf{0.60} \\
 &  &  &  &  &  & 4 & $140\pm20$ & 0.54 & 2.4 & 391 & 0.27 \\
 &  &  &  &  &  & \quad5 E & $520\pm60$ & 1.31 & 3.4 & \bf{252} & 0.12 \\
KOI-701 & 6 & 18.0 & 7.4 & 4.04 & 0.01 & 1 & $8.4\pm0.8$ & 0.07 & 0.6 & 621 & \bf{0.87} \\
 &  &  &  &  &  & 2 & $26.6\pm2.6$ & 0.15 & 0.8 & 423 & 0.52 \\
 &  &  &  &  &  & 3 & $39.1\pm3.7$ & 0.19 & 0.9 & 372 & 0.41 \\
 &  &  &  &  &  & 4 & $57\pm6$ & 0.25 & 1.0 & 327 & 0.33 \\
 &  &  &  &  &  & 5 & $84\pm8$ & 0.32 & 1.1 & \bf{288} & 0.25 \\
 &  &  &  &  &  & 6 & $180\pm20$ & 0.54 & 1.3 & \bf{223} & 0.15 \\
 &  &  &  &  &  & \quad7 E & $390\pm40$ & 0.90 & 1.6 & 172 & 0.09 \\
KOI-1306 & 1 & 5.7 & 0.4 & 4.12 & 0.62 & 1 & $13.7\pm2.1$ & 0.11 & 1.5 & 756 & \bf{0.85} \\
 &  &  &  &  &  & \quad2 E & $55\pm9$ & 0.28 & 2.1 & 476 & 0.43 \\
KOI-2722 & 0 & - & - & 0.54 & - & \quad 1 E & $23.4\pm1.5$ & 0.17 & 1.4 & 774 & \bf{0.78} \\
 &  &  &  &  &  & \quad 2 E & $33.0\pm1.8$ & 0.21 & 1.5 & 690 & \bf{0.67} \\
 &  &  &  &  &  & \quad 3 E & $46.5\pm2.3$ & 0.27 & 1.6 & 615 & \bf{0.56} \\
KOI-1358 & 0 & - & - & 0.01 & - & \quad 1 E & $13.6\pm1.0$ & 0.10 & 1.6 & 522 & \bf{0.74} \\
 &  &  &  &  &  & \quad 2 E & $21.0\pm1.3$ & 0.14 & 1.8 & 451 & \bf{0.60} \\
KOI-1627 & 0 & - & - & 0.24 & - & \quad 1 E & $16.6\pm1.2$ & 0.13 & 1.9 & 586 & \bf{0.73} \\
 &  &  &  &  &  & \quad 2 E & $27.4\pm1.5$ & 0.17 & 2.1 & 497 & \bf{0.57} \\
KOI-1833 & 0 & - & - & 0.74 & - & \quad 1 E & $11.3\pm0.6$ & 0.08 & 2.0 & 514 & \bf{0.72} \\
 &  &  &  &  &  & \quad 2 E & $16.4\pm0.7$ & 0.10 & 2.2 & 455 & \bf{0.60} \\
KOI-3158 & 0 & - & - & 0.23 & - & \quad 1 E & $12.7\pm0.6$ & 0.09 & 0.4 & 566 & \bf{0.71} \\
 &  &  &  &  &  & \quad 2 E & $16.4\pm0.7$ & 0.11 & 0.4 & 520 & \bf{0.62} \\
 &  &  &  &  &  & \quad 3 E & $21.1\pm0.8$ & 0.13 & 0.5 & 478 & \bf{0.55} \\
KOI-2055 & 0 & - & - & 0.29 & - & \quad 1 E & $13.6\pm1.0$ & 0.11 & 1.3 & 703 & \bf{0.70} \\
 &  &  &  &  &  & \quad 2 E & $20.6\pm1.2$ & 0.14 & 1.5 & 612 & \bf{0.56} \\
KOI-245 & 3 & 1.0 & 1.3 & 1.56 & 0.17 & 1 & $16.8\pm0.9$ & 0.12 & 0.3 & 582 & \bf{0.70} \\
 &  &  &  &  &  & 2 & $26.1\pm1.4$ & 0.16 & 0.3 & 502 & \bf{0.56} \\
 &  &  &  &  &  & 3 & $32.6\pm1.7$ & 0.18 & 0.3 & 467 & 0.50 \\
 &  &  &  &  &  & \quad4 E & $63.1\pm3.3$ & 0.28 & 0.4 & 374 & 0.33 \\
KOI-749 & 0 & - & - & 0.49 & - & \quad 1 E & $11.4\pm0.6$ & 0.10 & 1.5 & 711 & \bf{0.69} \\
 &  &  &  &  &  & \quad 2 E & $16.4\pm0.7$ & 0.12 & 1.7 & 630 & \bf{0.57} \\
KOI-730 & 0 & - & - & 0.33 & - & \quad 1 E & $27.9\pm1.5$ & 0.18 & 2.3 & 620 & \bf{0.69} \\
 &  &  &  &  &  & \quad 2 E & $39.0\pm1.8$ & 0.22 & 2.5 & 554 & \bf{0.58} \\
KOI-719 & 1 & 1.1 & 1.5 & 1.36 & 0.66 & 1 & $14.8\pm2.0$ & 0.11 & 0.8 & 514 & \bf{0.69} \\
 &  &  &  &  &  & \quad2 E & $88\pm12$ & 0.35 & 1.2 & \bf{284} & 0.24 \\
KOI-1060 & 0 & - & - & 0.22 & - & \quad 1 E & $32.7\pm2.6$ & 0.22 & 2.1 & 703 & \bf{0.68} \\
 &  &  &  &  &  & \quad 2 E & $52.7\pm3.5$ & 0.30 & 2.3 & 599 & 0.53 \\
KOI-3083 & 0 & - & - & 0.56 & - & \quad 1 E & $13.2\pm0.5$ & 0.11 & 0.7 & 751 & \bf{0.66} \\
 &  &  &  &  &  & \quad 2 E & $16.9\pm0.5$ & 0.13 & 0.7 & 692 & \bf{0.57} \\
KOI-156 & 0 & - & - & 0.10 & - & \quad 1 E & $17.9\pm1.0$ & 0.12 & 1.6 & 476 & \bf{0.61} \\
KOI-137 & 0 & - & - & 0.13 & - & \quad 1 E & $31.2\pm3.1$ & 0.19 & 3.1 & 539 & \bf{0.60} \\
KOI-1151 & 0 & - & - & 0.85 & - & \quad 1 E & $33.0\pm2.2$ & 0.20 & 1.0 & 564 & \bf{0.59} \\
KOI-1015 & 0 & - & - & 0.62 & - & \quad 1 E & $36.1\pm3.5$ & 0.22 & 2.3 & 590 & \bf{0.58} \\
KOI-2029 & 0 & - & - & 0.31 & - & \quad 1 E & $23.7\pm1.6$ & 0.15 & 0.9 & 514 & \bf{0.56} \\
KOI-664 & 0 & - & - & 0.06 & - & \quad 1 E & $40.3\pm3.0$ & 0.23 & 1.5 & 618 & \bf{0.56} \\
KOI-2693 & 0 & - & - & 0.01 & - & \quad 1 E & $19.1\pm1.4$ & 0.12 & 1.0 & 430 & 0.53 \\
KOI-1590 & 0 & - & - & 0.64 & - & \quad 1 E & $28.7\pm3.3$ & 0.16 & 1.8 & 452 & 0.53 \\
KOI-279 & 0 & - & - & 0.13 & - & \quad 1 E & $56\pm6$ & 0.30 & 1.3 & 586 & 0.53 \\
KOI-1930 & 0 & - & - & 0.60 & - & \quad 1 E & $72\pm7$ & 0.35 & 2.3 & 541 & 0.52 \\
KOI-70 & 1 & 7.4 & 2.9 & 3.51 & 0.42 & 1 & $39.1\pm5.4$ & 0.22 & 1.2 & 498 & 0.52 \\
 &  &  &  &  &  & \quad2 E & $130\pm20$ & 0.49 & 1.7 & 331 & 0.24 \\
KOI-720 & 0 & - & - & 0.14 & - & \quad 1 E & $34.8\pm3.6$ & 0.20 & 2.9 & 477 & 0.51 \\
KOI-1860 & 0 & - & - & 0.02 & - & \quad 1 E & $49.5\pm5.6$ & 0.27 & 1.7 & 512 & 0.49 \\
KOI-1475 & 0 & - & - & 0.94 & - & \quad 1 E & $24.6\pm3.0$ & 0.14 & 1.7 & 377 & 0.48 \\
KOI-1194 & 0 & - & - & 0.52 & - & \quad 1 E & $29.0\pm2.5$ & 0.16 & 1.8 & 374 & 0.47 \\
KOI-2025 & 0 & - & - & 0.21 & - & \quad 1 E & $40.5\pm2.4$ & 0.24 & 2.3 & 647 & 0.47 \\
KOI-733 & 0 & - & - & 0.22 & - & \quad 1 E & $35.5\pm3.5$ & 0.20 & 2.8 & 437 & 0.46 \\
KOI-2169 & 0 & - & - & 0.87 & - & \quad 1 E & $7.6\pm0.4$ & 0.07 & 0.7 & 868 & 0.46 \\
KOI-2163 & 0 & - & - & 0.06 & - & \quad 1 E & $46.3\pm3.1$ & 0.26 & 1.7 & 532 & 0.44 \\
KOI-3319 & 0 & - & - & 0.01 & - & \quad 1 E & $45.9\pm5.3$ & 0.25 & 2.1 & 517 & 0.44 \\
KOI-2352 & 0 & - & - & 0.28 & - & \quad 1 E & $20.3\pm1.2$ & 0.16 & 1.1 & 845 & 0.44 \\
KOI-1681 & 0 & - & - & 0.15 & - & \quad 1 E & $12.7\pm1.1$ & 0.09 & 1.3 & 415 & 0.44 \\
KOI-1413 & 0 & - & - & 0.14 & - & \quad 1 E & $56.2\pm3.8$ & 0.28 & 1.8 & 458 & 0.44 \\
KOI-2597 & 0 & - & - & 0.13 & - & \quad 1 E & $17.7\pm1.0$ & 0.14 & 1.8 & 791 & 0.43 \\
KOI-2220 & 0 & - & - & 0.19 & - & \quad 1 E & $19.1\pm1.6$ & 0.14 & 1.2 & 695 & 0.42 \\
KOI-1161 & 0 & - & - & 0.18 & - & \quad 1 E & $21.4\pm1.9$ & 0.14 & 2.1 & 574 & 0.42 \\
KOI-582 & 0 & - & - & 0.08 & - & \quad 1 E & $30.3\pm2.3$ & 0.18 & 1.8 & 466 & 0.41 \\
KOI-82 & 0 & - & - & 0.92 & - & \quad 1 E & $38.3\pm2.9$ & 0.21 & 0.9 & 408 & 0.41 \\
KOI-157 & 1 & 3.7 & 6.9 & 3.15 & 0.69 & 1 & $75\pm8$ & 0.34 & 2.7 & 439 & 0.41 \\
 &  &  &  &  &  & \quad2 E & $170\pm20$ & 0.60 & 3.4 & 334 & 0.24 \\
KOI-864 & 0 & - & - & 0.09 & - & \quad 1 E & $44.0\pm4.6$ & 0.24 & 2.9 & 453 & 0.40 \\
KOI-939 & 0 & - & - & 0.24 & - & \quad 1 E & $20.3\pm2.1$ & 0.15 & 1.9 & 640 & 0.40 \\
KOI-898 & 0 & - & - & 0.08 & - & \quad 1 E & $39.1\pm3.6$ & 0.20 & 2.8 & 360 & 0.40 \\
KOI-841 & 2 & 7.1 & 0.4 & 4.35 & 0.15 & 1 & $63\pm11$ & 0.31 & 2.7 & 409 & 0.39 \\
 &  &  &  &  &  & 2 & $130\pm30$ & 0.51 & 3.3 & 320 & 0.25 \\
 &  &  &  &  &  & \quad3 E & $580\pm100$ & 1.35 & 4.7 & 196 & 0.09 \\
KOI-408 & 0 & - & - & 0.51 & - & \quad 1 E & $59\pm7$ & 0.29 & 2.6 & 461 & 0.39 \\
KOI-1909 & 0 & - & - & 0.26 & - & \quad 1 E & $55\pm6$ & 0.29 & 1.9 & 500 & 0.38 \\
KOI-2715 & 0 & - & - & 0.62 & - & \quad 1 E & $26.1\pm2.9$ & 0.14 & 3.9 & 379 & 0.38 \\
KOI-1278 & 0 & - & - & 0.52 & - & \quad 1 E & $73\pm7$ & 0.35 & 2.0 & 455 & 0.38 \\
KOI-1867 & 0 & - & - & 0.53 & - & \quad 1 E & $31.3\pm3.6$ & 0.16 & 1.7 & 323 & 0.37 \\
KOI-899 & 0 & - & - & 0.01 & - & \quad 1 E & $33.1\pm3.5$ & 0.16 & 1.9 & \bf{293} & 0.37 \\
KOI-1589 & 0 & - & - & 0.56 & - & \quad 1 E & $82\pm9$ & 0.37 & 2.4 & 440 & 0.37 \\
KOI-884 & 0 & - & - & 0.45 & - & \quad 1 E & $53\pm7$ & 0.25 & 2.9 & 362 & 0.37 \\
KOI-829 & 0 & - & - & 0.07 & - & \quad 1 E & $76\pm8$ & 0.36 & 3.5 & 442 & 0.37 \\
KOI-94 & 0 & - & - & 0.19 & - & \quad 1 E & $130\pm20$ & 0.55 & 4.2 & 452 & 0.36 \\
KOI-2038 & 0 & - & - & 0.11 & - & \quad 1 E & $37.0\pm2.3$ & 0.21 & 1.8 & 494 & 0.36 \\
KOI-1557 & 0 & - & - & 0.58 & - & \quad 1 E & $18.2\pm1.9$ & 0.12 & 2.0 & 499 & 0.35 \\
KOI-571 & 2 & 19.9 & 5.2 & 4.87 & 0.07 & 1 & $40.9\pm5.6$ & 0.19 & 0.8 & \bf{294} & 0.35 \\
 &  &  &  &  &  & 2 & $73\pm10$ & 0.28 & 0.9 & \bf{242} & 0.24 \\
 &  &  &  &  &  & \quad3 E & $230\pm40$ & 0.61 & 1.2 & 164 & 0.11 \\
KOI-1905 & 0 & - & - & 0.01 & - & \quad 1 E & $72\pm8$ & 0.32 & 1.7 & 374 & 0.34 \\
KOI-116 & 0 & - & - & 0.34 & - & \quad 1 E & $86\pm10$ & 0.38 & 1.3 & 425 & 0.34 \\
KOI-2732 & 0 & - & - & 0.22 & - & \quad 1 E & $100\pm20$ & 0.44 & 1.3 & 426 & 0.34 \\
KOI-665 & 0 & - & - & 0.01 & - & \quad 1 E & $11.2\pm1.0$ & 0.10 & 1.5 & 893 & 0.34 \\
KOI-1931 & 0 & - & - & 0.20 & - & \quad 1 E & $15.2\pm0.8$ & 0.12 & 1.5 & 661 & 0.33 \\
KOI-886 & 0 & - & - & 0.46 & - & \quad 1 E & $33.2\pm2.2$ & 0.16 & 1.6 & \bf{298} & 0.32 \\
KOI-1432 & 0 & - & - & 0.07 & - & \quad 1 E & $87\pm13$ & 0.38 & 2.0 & 397 & 0.32 \\
KOI-945 & 0 & - & - & 0.04 & - & \quad 1 E & $107\pm7$ & 0.46 & 2.6 & 424 & 0.32 \\
KOI-869 & 0 & - & - & 0.08 & - & \quad 1 E & $84\pm12$ & 0.35 & 3.8 & 349 & 0.31 \\
KOI-111 & 0 & - & - & 0.03 & - & \quad 1 E & $110\pm20$ & 0.42 & 2.8 & 376 & 0.30 \\
KOI-1364 & 0 & - & - & 0.41 & - & \quad 1 E & $34.9\pm3.3$ & 0.20 & 3.0 & 508 & 0.30 \\
KOI-1832 & 0 & - & - & 0.03 & - & \quad 1 E & $110\pm20$ & 0.44 & 3.6 & 381 & 0.30 \\
KOI-658 & 0 & - & - & 0.60 & - & \quad 1 E & $20.7\pm1.8$ & 0.15 & 1.4 & 649 & 0.30 \\
KOI-1895 & 0 & - & - & 0.11 & - & \quad 1 E & $64\pm6$ & 0.27 & 2.6 & \bf{289} & 0.30 \\
KOI-2926 & 0 & - & - & 0.49 & - & \quad 1 E & $73\pm8$ & 0.27 & 2.8 & \bf{260} & 0.29 \\
KOI-1647 & 0 & - & - & 0.07 & - & \quad 1 E & $74\pm9$ & 0.34 & 2.0 & 460 & 0.29 \\
KOI-941 & 0 & - & - & 0.36 & - & \quad 1 E & $75\pm12$ & 0.32 & 4.7 & 343 & 0.29 \\
KOI-3741 & 0 & - & - & 0.01 & - & \quad 1 E & $35.3\pm3.1$ & 0.21 & 2.2 & 709 & 0.28 \\
KOI-700 & 0 & - & - & 0.91 & - & \quad 1 E & $120\pm20$ & 0.48 & 2.1 & 388 & 0.28 \\
KOI-1563 & 0 & - & - & 0.62 & - & \quad 1 E & $26.9\pm2.5$ & 0.17 & 3.6 & 491 & 0.28 \\
KOI-2135 & 0 & - & - & 0.12 & - & \quad 1 E & $140\pm20$ & 0.53 & 1.9 & 414 & 0.28 \\
KOI-2433 & 0 & - & - & 0.55 & - & \quad 1 E & $150\pm20$ & 0.57 & 2.9 & 395 & 0.28 \\
KOI-2086 & 0 & - & - & 0.33 & - & \quad 1 E & $15.2\pm0.6$ & 0.13 & 2.7 & 933 & 0.27 \\
KOI-1102 & 0 & - & - & 0.75 & - & \quad 1 E & $32.4\pm2.7$ & 0.20 & 2.8 & 630 & 0.27 \\
KOI-3097 & 0 & - & - & 0.31 & - & \quad 1 E & $15.6\pm0.6$ & 0.13 & 1.3 & 1033 & 0.27 \\
KOI-1445 & 0 & - & - & 0.01 & - & \quad 1 E & $150\pm30$ & 0.59 & 1.4 & 391 & 0.27 \\
KOI-232 & 0 & - & - & 0.93 & - & \quad 1 E & $110\pm20$ & 0.46 & 2.1 & 391 & 0.26 \\
KOI-520 & 0 & - & - & 0.18 & - & \quad 1 E & $110\pm20$ & 0.42 & 1.8 & 314 & 0.26 \\
KOI-2707 & 0 & - & - & 0.29 & - & \quad 1 E & $110\pm20$ & 0.45 & 2.3 & 335 & 0.25 \\
KOI-152 & 0 & - & - & 0.64 & - & \quad 1 E & $160\pm20$ & 0.60 & 3.6 & 381 & 0.25 \\
KOI-1332 & 0 & - & - & 0.03 & - & \quad 1 E & $170\pm30$ & 0.63 & 3.8 & 337 & 0.25 \\
KOI-2485 & 0 & - & - & 0.17 & - & \quad 1 E & $16.4\pm1.2$ & 0.12 & 1.8 & 575 & 0.25 \\
KOI-877 & 0 & - & - & 0.32 & - & \quad 1 E & $40.0\pm3.4$ & 0.20 & 1.6 & 343 & 0.24 \\
KOI-775 & 0 & - & - & 0.04 & - & \quad 1 E & $78\pm9$ & 0.30 & 2.4 & \bf{253} & 0.24 \\
KOI-757 & 0 & - & - & 0.01 & - & \quad 1 E & $110\pm20$ & 0.41 & 4.2 & \bf{295} & 0.24 \\
KOI-510 & 0 & - & - & 0.05 & - & \quad 1 E & $78\pm11$ & 0.34 & 3.5 & 413 & 0.24 \\
KOI-117 & 0 & - & - & 0.42 & - & \quad 1 E & $23.3\pm2.0$ & 0.17 & 1.4 & 692 & 0.23 \\
KOI-935 & 0 & - & - & 0.03 & - & \quad 1 E & $180\pm30$ & 0.68 & 4.1 & 374 & 0.23 \\
KOI-285 & 0 & - & - & 0.04 & - & \quad 1 E & $94\pm9$ & 0.43 & 2.2 & 490 & 0.22 \\
KOI-671 & 0 & - & - & 0.62 & - & \quad 1 E & $26.4\pm2.0$ & 0.17 & 1.5 & 614 & 0.22 \\
KOI-834 & 0 & - & - & 0.84 & - & \quad 1 E & $120\pm20$ & 0.47 & 2.5 & 351 & 0.22 \\
KOI-509 & 0 & - & - & 0.22 & - & \quad 1 E & $120\pm20$ & 0.46 & 2.5 & 334 & 0.22 \\
KOI-904 & 0 & - & - & 0.93 & - & \quad 1 E & $100\pm20$ & 0.37 & 2.2 & \bf{252} & 0.22 \\
KOI-723 & 0 & - & - & 0.04 & - & \quad 1 E & $74\pm10$ & 0.33 & 4.2 & 357 & 0.22 \\
KOI-1436 & 0 & - & - & 0.19 & - & \quad 1 E & $31.3\pm3.6$ & 0.19 & 2.3 & 555 & 0.21 \\
KOI-435 & 2 & 1.8 & 0.3 & 6.69 & 0.84 & 1 & $160\pm30$ & 0.55 & 2.3 & 321 & 0.21 \\
 &  &  &  &  &  & 2 & $320\pm60$ & 0.90 & 2.8 & \bf{252} & 0.13 \\
 &  &  &  &  &  & \quad3 E & $1380\pm250$ & 2.35 & 4.0 & 155 & 0.05 \\
KOI-2073 & 0 & - & - & 0.08 & - & \quad 1 E & $130\pm20$ & 0.46 & 2.8 & \bf{274} & 0.20 \\
KOI-812 & 0 & - & - & 0.04 & - & \quad 1 E & $110\pm20$ & 0.38 & 2.5 & \bf{224} & 0.20 \\
KOI-474 & 0 & - & - & 0.54 & - & \quad 1 E & $230\pm40$ & 0.77 & 3.8 & 336 & 0.20 \\
KOI-907 & 0 & - & - & 0.91 & - & \quad 1 E & $250\pm50$ & 0.76 & 4.5 & 309 & 0.19 \\
KOI-1422 & 0 & - & - & 0.04 & - & \quad 1 E & $110\pm20$ & 0.37 & 1.9 & 198 & 0.19 \\
KOI-710 & 0 & - & - & 0.69 & - & \quad 1 E & $12.5\pm0.7$ & 0.12 & 1.7 & 998 & 0.19 \\
KOI-623 & 0 & - & - & 0.40 & - & \quad 1 E & $42.2\pm3.5$ & 0.23 & 1.3 & 592 & 0.19 \\
KOI-282 & 0 & - & - & 0.01 & - & \quad 1 E & $280\pm50$ & 0.83 & 1.5 & \bf{299} & 0.19 \\
KOI-620 & 0 & - & - & 0.84 & - & \quad 1 E & $230\pm20$ & 0.74 & 7.6 & \bf{298} & 0.18 \\
KOI-3925 & 0 & - & - & 0.38 & - & \quad 1 E & $17.9\pm1.6$ & 0.13 & 3.5 & 779 & 0.18 \\
KOI-2167 & 0 & - & - & 0.05 & - & \quad 1 E & $260\pm50$ & 0.82 & 1.8 & 303 & 0.17 \\
KOI-1426 & 0 & - & - & 0.02 & - & \quad 1 E & $290\pm30$ & 0.88 & 5.3 & \bf{294} & 0.16 \\
KOI-1127 & 0 & - & - & 0.81 & - & \quad 1 E & $14.2\pm1.1$ & 0.12 & 2.0 & 681 & 0.16 \\
KOI-191 & 0 & - & - & 0.99 & - & \quad 1 E & $180\pm40$ & 0.61 & 3.4 & 301 & 0.16 \\
KOI-1430 & 0 & - & - & 0.98 & - & \quad 1 E & $200\pm30$ & 0.57 & 3.2 & \bf{215} & 0.15 \\
KOI-806 & 0 & - & - & 0.16 & - & \quad 1 E & $310\pm40$ & 0.89 & 2.5 & \bf{247} & 0.15 \\
KOI-612 & 0 & - & - & 0.10 & - & \quad 1 E & $290\pm40$ & 0.79 & 3.6 & \bf{235} & 0.15 \\
KOI-481 & 0 & - & - & 0.02 & - & \quad 1 E & $160\pm40$ & 0.56 & 4.0 & \bf{282} & 0.14 \\
KOI-2714 & 0 & - & - & 0.09 & - & \quad 1 E & $640\pm120$ & 1.55 & 3.4 & \bf{272} & 0.13 \\
KOI-1258 & 0 & - & - & 0.96 & - & \quad 1 E & $430\pm70$ & 1.11 & 5.1 & \bf{218} & 0.10 \\
KOI-564 & 0 & - & - & 0.77 & - & \quad 1 E & $530\pm110$ & 1.28 & 4.3 & \bf{222} & 0.10 \\
KOI-1922 & 0 & - & - & 0.02 & - & \quad 1 E & $790\pm220$ & 1.69 & 2.9 & 199 & 0.08 \\
KOI-2183 & 0 & - & - & 0.70 & - & \quad 1 E & $770\pm190$ & 1.64 & 3.0 & 188 & 0.07 \\
KOI-518 & 0 & - & - & 0.84 & - & \quad 1 E & $940\pm190$ & 1.59 & 3.1 & 124 & 0.05 \\
KOI-2842 & 0 & - & - & 0.27 & - & \quad 1 E & $9.5\pm0.8$ & 0.07 & 3.2 & 559 & 0.00 \\

\end{longtable}
\label{tab:allpredictions}
\label{lastpage}
\end{ThreePartTable}
\twocolumn

\end{document}